\documentclass[%
 reprint,
showpacs,preprintnumbers,
 amsmath,amssymb,
 aps,
pra,
]{revtex4-1}
\usepackage{dcolumn}% Align table columns on decimal point
\usepackage{bm}% bold math

\usepackage{graphicx}
\usepackage{epstopdf}
\usepackage{float}
\usepackage{subfigure}
\usepackage{algorithm}
\usepackage{algpseudocode}
\usepackage{caption}

\usepackage{subfig}

\newtheorem{theorem}{Theorem}

\begin{document}

%\preprint{APS/123-QED}

\title{Quantum Algorithm for DOA Estimation in Hybrid Massive MIMO}

\author{Fan-Xu Meng}
\affiliation{National Mobile Communications Research Laboratory, Southeast University, Nanjing 211189, China}
\affiliation{Frontiers Science Center for Mobile Information Communication and Security, Nanjing, Jiangsu 211111, China}
\affiliation{Quantum Information Center of Southeast University, Nanjing 211189, China}
\author{Xu-Tao Yu}
\email{yuxutao@seu.edu.cn}
\affiliation{State Key Lab of Millimeter Waves, Southeast University, Nanjing 211189, China}
\affiliation{Frontiers Science Center for Mobile Information Communication and Security, Nanjing, Jiangsu 211111, China}
\affiliation{Quantum Information Center of Southeast University, Nanjing 211189, China}
\author{Ze-Tong Li}
\affiliation{State Key Lab of Millimeter Waves, Southeast University, Nanjing 211189, China}
\affiliation{Frontiers Science Center for Mobile Information Communication and Security, Nanjing, Jiangsu 211111, China}
\affiliation{Quantum Information Center of Southeast University, Nanjing 211189, China}
%\author{Wei-Dong Li}
%\affiliation{State Key Lab of Millimeter Waves, Southeast University, Nanjing 211189, China}
\author{Zai-Chen Zhang}
\affiliation{National Mobile Communications Research Laboratory, Southeast University, Nanjing 211189, China}
\affiliation{Frontiers Science Center for Mobile Information Communication and Security, Nanjing, Jiangsu 211111, China}
\affiliation{Quantum Information Center of Southeast University, Nanjing 211189, China}
\affiliation{Purple Mountain Laboratories, Nanjing, Jiangsu 211111, China}

\date{\today}% It is always \today, today,

\begin{abstract}
The direction of arrival (DOA) estimation in array signal processing is an important research area. The effectiveness of the direction of arrival greatly determines the performance of multi-input multi-output (MIMO) antenna systems. The multiple signal classification (MUSIC) algorithm, which is the most canonical and widely used subspace-based method, has a moderate estimation performance of DOA. However, in hybrid
massive MIMO systems, the received signals at the antennas are not sent to the receiver directly, and spatial covariance matrix, which is essential in MUSIC algorithm, is thus unavailable. 
Therefore, the spatial covariance matrix reconstruction is required for the application of MUSIC in hybrid massive MIMO systems. In this article, we present a quantum algorithm for MUSIC-based DOA estimation in hybrid massive MIMO systems. Compared with the best-known classical algorithm, our quantum algorithm can achieve an exponential
speedup on some parameters and a polynomial speedup on others under some mild conditions. In our scheme, we first present the quantum subroutine for the beam sweeping based spatial covariance matrix reconstruction, where we implement a quantum singular vector transition process to avoid extending the steering vectors matrix into the Hermitian form. Second, a variational quantum density matrix eigensolver (VQDME) is proposed for obtaining signal and noise subspaces, where we design a novel objective function in the form of the trace of density matrices product. Finally, a quantum labeling operation is proposed for the direction of arrival estimation of the signal.
\end{abstract}
%\keywords{Throughput capacity, transport capacity, quantum network.}%Use showkeys class option if keyword
\maketitle

\section{Introduction}
The direction of space signal arrival estimation \cite{1} is a crucial issue in signal processing, which has also gained significant attention in many applications, including smart antenna systems and wireless locations. The core of the direction of arrival estimation is to determine signal source location in a certain space area where the entire spatial spectrum consists of the target, observation, and estimation stages \cite{2}. As the signals are assumed to be distributed in the space of all the directions, the spatial spectrum of the signal can be exploited to provide a good effect for the direction of arrival estimation. Spatial spectrum-based algorithms are the most successful algorithmic frameworks in the past few decades. Among these are the famous estimation of signal parameters via rotational invariance technique (ESPRIT) and multiple signal classification (MUSIC). ESPRIT \cite{3,4,5,6,7} has exploited the invariance property for direction estimation. However, these algorithms \cite{3,4,5,6,7} have almost the same performance and can only be applied to array structures with some peculiar geometries. MUSIC was first proposed by Schmidt in \cite{8}. It is the most widely used subspace and spectral estimation method based on the decomposition of eigenvalues. Moreover, MUSIC algorithm can also match some types of irregularly spaced arrays.

Compared with conventional MIMO, Massive MIMO, which is one of the most important enabling technologies in the fifth generation (5G) systems \cite{9}, can obtain significant array gain with a large number of antennas. Moreover, the frequency resources at the millimeterwave system can be used efficiently with massive MIMO. To reduce the cost of radio frequency chains at millimeterwave bands, the hybrid massive MIMO structure \cite{10,11,12,13} have been proposed, where received signals are fed to the analog phase shifters and then combined in the analog domain before being sent to the receiver. Therefore, the receiver cannot directly obtain the received signals at the antennas, which results in the failure of constructing the spatial covariance matrix in DOA estimation. For the application of MUSIC in the hybrid massive MIMO, a novel technique for spatial covariance matrix reconstruction \cite{14} was designed. In this approach, analog beamformer switches the beam direction to predetermined DOA angles in turn, and the average of received power in each sweeping beam is only required. Thus, the spatial covariance matrix can be reconstructed by solving the regularized least square problem, which can ensure that MUSIC can be appropriate for hybrid massive MIMO systems. In this technique, the reconstruction of the spatial covariance matrix can be implemented in the runtime ${\rm O}\left( {M^\omega   +  M^2 Q} \right)$, where $\omega  \in \left( {{\rm{5,6}}} \right)$, $M$ and $Q$ are the number of antenna elements and predetermined DOA angles, respectively. Subsequently, based on the spatial covariance matrix, MUSIC algorithm can be performed with the complexity ${\rm O}\left( {M^3  +  M^2 K} \right)$, where $K$ is the size of the direction searching space. Therefore, MUSIC-based DOA estimation in hybrid massive MIMO system runs in approximate ${\rm O}\left( {\ M^\omega  }+M^2KQ \right)$ complexity. However, in the 6G and post 6G era, with the increase of the number of antenna elements, signal sources, and snapshots, high time and space complexity will be the limit and the drawback of the classical  algorithm for the hybrid massive MIMO systems.

Quantum computing was established as a promising extension to classical computation and was theoretically shown to perform significantly better in selected computational problems. We are witnessing the influence of quantum
information processing on its classical counterpart. Following the discovery of quantum algorithms for factoring \cite{15}, database searching \cite{16} and quantum matrix inverse \cite{17}, a range of quantum algorithms \cite{18,19,20,21,22,23,24,25,26} have shown the capability of outperforming classical algorithms in machine learning. However, only a very small number of research works focus on connecting quantum algorithms to wireless communication systems. Among these are quantum search algorithms assisted multi-user detection \cite{27}, quantum-inspired tabu search algorithm for antenna selection in massive MIMO \cite{28}, and quantum-assisted routing optimization \cite{29} and quantum-aided multi-user transmission \cite{30}. All of the aforementioned research makes full use of Grover algorithm and its variants to solve the search problems, which achieve quadratic speedup over classical counterparts. Moreover, the quantum algorithm for MUSIC \cite{31} has also been proposed with the polynomial speedup. However, above these algorithms will require the enormous number of qubits, quantum gates and circuit structures with deep depth.

Fortunately, noisy intermediate-scale quantum (NISQ) devices \cite{32} are considered as a significant step toward more powerful quantum computer and shown quantum supremacy. Therefore, an important direction is to find useful algorithms that can work on NISQ devices. The leading strategy for various problems using NISQ devices are called variational quantum algorithms (VQA) \cite{33}, which can be implemented in a shallow-depth parameterized quantum circuit. These parameters will be optimized in classical computers with respect to certain loss functions. Recently, a number of variational quantum algorithms have been proposed, including the ground and excited states preparation of Hamiltonian or density matrix \cite{34,35,36,37,38}, singular value decomposition \cite{39,40}, matrix operations \cite{41,42,43}, quantum state fidelity estimation \cite{44}, quantum Gibbs state preparation \cite{45,46} and the calculation of the Green’s function \cite{47}. Furthermore, unlike the strong need for error correction in fault-tolerant quantum computation, the noise in shallow quantum circuits can be suppressed via quantum error mitigation \cite{48,49,50}, which can demonstrate the feasibility of quantum computing with NISQ devices.

In the present study, we propose a quantum algorithm for MUSIC-based DOA estimation in the hybrid massive MIMO system. Our work consists of three contributions. First, we present the quantum subroutine for the reconstruction of the spatial covariance matrix, where the efficient quantum circuit is designed to prepare any row vector of the steering vectors matrix, and propose the quantum singular vector transition process to avoid extending the steering vectors matrix into a Hermitian form. Second, we design the variational quantum density matrix eigensolver (VQDME) algorithm for the eigen-decomposition part of MUSIC algorithm, where orthogonal base vectors of signal subspace and noise subspace are obtained, effectively. Third, a quantum labeling  algorithm is presented for the direction estimation. Consequently, the time complexity of our quantum algorithm is analyzed, and an exponential
speedup on some parameters and a polynomial speedup on others compared with classical counterparts are shown under some mild conditions.

The remainder of this paper is organized as follows. In Sec.\ref{sec_Preliminaries}, we provide some requisite background information and give a brief overview of the reconstruction of the spatial covariance matrix in hybrid massive MIMO systems and the classical MUSIC algorithm. In Sec.\ref{sec_mathematical description}, the key quantum subroutines are introduced in detail. Finally, a summary and discussions are included in Sec.\ref{sec_Conclusion}.

\section{Preliminaries}
\label{sec_Preliminaries}
In this section, we provide the necessary background to better understand this paper. In Sec.\ref{sec_Preliminaries} A, we briefly view some basic notations used in this paper. In Sec.\ref{sec_Preliminaries} B, a detailed overview of the reconstruction of the spatial covariance matrix algorithm is presented. Then, a detailed overview of MUSIC algorithms is depicted in Sec.\ref{sec_Preliminaries} C.

\subsection{Notation}
Throughout, we obey the following conventions. $\mathop I\nolimits_N $ refers to an $N \times N$ identity matrix. For an arbitrary matrix $A \in C^{M \times N}$, let $A = U\Lambda  V^H  $ be the singular value decomposition (SVD) of $A$. Here, $U = \left[ {\mathop u\nolimits_1 ,\mathop u\nolimits_2  \ldots \mathop u\nolimits_M } \right]$ and $V = \left[ {\mathop v\nolimits_1 ,\mathop v\nolimits_2  \ldots \mathop v\nolimits_N } \right]$ are unitary matrices with the left and right singular vectors of $X$ as columns, respectively.
$\Lambda $ is a diagonal matrix with a singular value ${\mathop \sigma \nolimits_i }$ as a diagonal element. The economy-sized SVD can be $A = \sum\limits_{i = 1}^r {\mathop \sigma \nolimits_i \mathop u\nolimits_i \mathop v\nolimits_i^H } $, where $r$ is the rank of $A$. The form of eigenvalue decomposition of $A A^H $ can be expressed as $AA^H {\rm{ = }}\sum\limits_{i = 1}^r {\mathop \sigma \nolimits_i^2 \mathop u\nolimits_i \mathop u\nolimits_i^H } $, where ${\mathop \sigma \nolimits_i^2 }$ is the eigenvalue of $A A^H $. $\left \| \cdot  \right \| _{F} $ refers to Frobenius norm, and $ \otimes $ denotes Kronecker product. The vectorization of a matrix $A$ can be denoted as ${\rm{vec}}\left( A \right) = \left( x_{11} ,x_{21}  \ldots x_{M1} ,x_{12} ,x_{22}  \ldots x_{M2} , \ldots ,x_{MN}  \right)^T  $ and ${\rm{vec}}\left( A \right) = \sum\limits_{i = 1}^r {\mathop \sigma \nolimits_i \mathop u\nolimits_i  \otimes \mathop v\nolimits_i } $.

%In this paper, we consider $A$ as a  Hermite matrix without loss of generality.
\subsection{REVIEW OF THE SPATIAL COVARIANCE MATRIX RECONSTRUCTION}
\label{sec_classcialMatrix}
Without loss of generality, we assume that there exists a set of predetermined DOA angles $\left\{ \theta^{\left( i \right)} \right\}_{i=1}^{Q} ,{\kern 1pt} {\kern 1pt}{\kern 1pt}\theta ^{\left( 1 \right)}  < \theta ^{\left( 2 \right)}  <  \ldots  < \theta^{\left( Q \right)} $. The beam direction is switched by the analog beamformer to the predetermined DOA angle in turn. Then, the combination of the received signals in the receiver can be represented by $c_q \left( t \right) = a^H \left( {\theta^{\left( q \right)} } \right)y\left( t \right)$, which is shown in Fig. 1.
\begin{figure}
\includegraphics[width=3.5in]{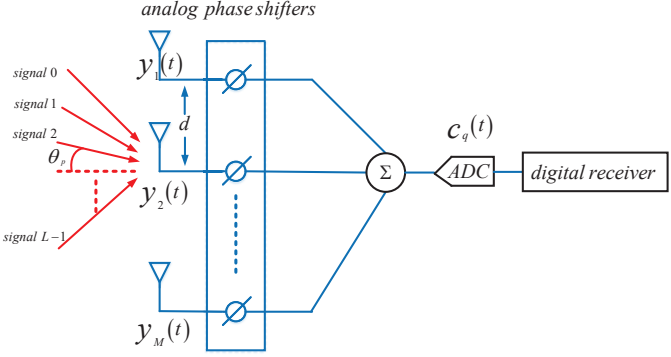}
\caption{Signal combination is sampled via analog-to-digital converter (ADC) before sending to the digital receiver.}
\label{fig_chain1111122}
\end{figure}
Thus, a sample of the signal combination can be given by
\begin{equation}
c_q \left[ n \right] = c_q \left( {nT_s } \right) = a^H \left( {\theta^q } \right)y\left[ n \right]
\end{equation}
where ${ T_s }$ denotes the sample period and $y\left[ n \right] = y\left( {n\mathop T\nolimits_s } \right)$. The steering vector $a\left( {\theta^{\left( q \right)} } \right)$ is defined as follows
\begin{equation}
a\left( {{\theta ^{\left( q \right)}}} \right) = {\left[ {1,{e^{ - j\frac{{2\pi d}}{\lambda }\sin \left( {{\theta ^{\left( q \right)}}} \right)}}, \ldots ,{e^{ - j\frac{{2\pi d\left( {M - 1} \right)}}{\lambda }\sin \left( {{\theta ^{\left( q \right)}}} \right)}}} \right]^T}
\end{equation}
where the distance $d$ between consecutive elements is equal to half of the signal wavelength $\lambda $. Let $P_q $ be the average power of $c\left[ n \right]$, then we can obtain 
\begin{equation}
\begin{array}{l}
P_q  = \frac{1}{N}\sum_{n = 0}^{N - 1} {{\left| {c_q \left[ n \right]} \right|}^2 } \\
 = a^H \left( {\theta ^{\left( q \right)} } \right)\frac{1}{N}\sum_{n = 0}^{N - 1} {y\left[ n \right]y^H \left[ n \right]} a\left( {\theta ^{\left( q \right)} } \right)
\end{array}
\end{equation}
When the number of samples is large enough, the sample average in Eq. (3) can be replaced by the statistical average, and we can obtain 
\begin{equation}
\mathop P\nolimits_q  = \mathop a\nolimits^H \left( {\mathop \theta \nolimits^{\left( q \right)} } \right)Ra\left( {\mathop \theta \nolimits^{\left( q \right)} } \right)
\end{equation}
Using the vectorization definition to Eq. (4), 
\begin{equation}
\begin{array}{l}
vec\left\{ {\mathop a\nolimits^H \left( {\mathop \theta \nolimits^{\left( q \right)} } \right)Ra\left( {\mathop \theta \nolimits^{\left( q \right)} } \right)} \right\}\\
 = \left[ {\mathop a\nolimits^T \left( {\mathop \theta \nolimits^{\left( q \right)} } \right) \otimes \mathop a\nolimits^H \left( {\mathop \theta \nolimits^{\left( q \right)} } \right)} \right]^Tvec\left( R \right)
\end{array}
\end{equation}
Let $a_q  = a\left( {\theta^{\left( q \right)} } \right)\otimes a^* \left( {\theta^{\left( q \right)} } \right)$ and $r = vec\left( R \right)$, then Eq. (4) can be rewritten as 
\begin{equation}
\mathop a\nolimits_q^T r = \mathop P\nolimits_q 
\end{equation}
Given the group of predetermined DOA angles, we can obtain 
\begin{equation}
Ar = P 
\end{equation}
where $A = {\left( {a_1 ,a_2 , \ldots a_Q } \right)}^T $ is a $Q \times M^2 $ matrix and  $P = {\left( {P_1 , P_2 , \ldots P_Q } \right)}^T $ is a $Q \times 1$ vector.
To avoid the ill-conditioned result, diagonal loading is adopted, and then we can obtain the vector form of the spatial covariance matrix as follows
\begin{equation}
\hat r = {\left( { A^H A + \sigma ^2  I_{M^2 } } \right)}^{ - 1} A^H P
\end{equation}
where $\hat R = unvec\left( {\hat r} \right)$ is the desired spatial covariance matrix.

\subsection{REVIEW OF THE CLASSICAL MUSIC ALGORITHM}
\label{sec_classcialMusic}

We assume that there are $L$ narrowband source signals incident upon an array of $M$ antenna elements. These $M$ elements are linearly spaced with equal distance between consecutive elements shown in Fig. 2.
\begin{figure}
\includegraphics[width=3.5in]{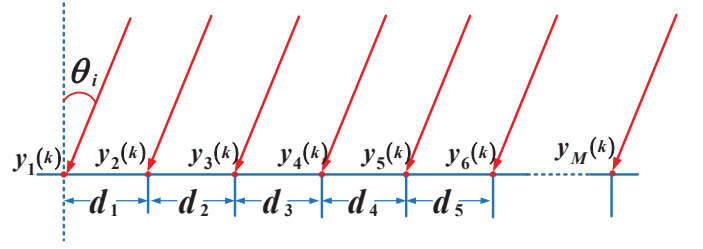}
\caption{One-dimensional structure of an antenna array.}
\label{fig_chain3344}
\end{figure}
Without loss of generality, we assume that the number of antenna elements is greater than the number of signals, i.e., $\left( {M \gg L} \right)$. The distance between consecutive elements is equal to half of the signal wavelength $\lambda $; namely, $d_1  = d_2  = d_3  =  \cdots  = d_M  = \frac{\lambda }{2}$.

Let $y\left( k \right) =  {\left[ { y_1 \left( k \right), y_2 \left( k \right), \ldots y_M \left( k \right)} \right]}^T $ be the $k$th observation data from $L$ source signals impinging on an array of $M$ elements; that is, 
\begin{equation}
y\left( k \right) = \sum\limits_{i = 0}^{L-1} {\mathop s\nolimits_i \left( k \right)a\left( {\mathop \theta \nolimits_i } \right)}  + n\left( k \right) = As\left( k \right) + n\left( k \right)
\end{equation}
where $A = \left[ {a\left( {{\theta _0}} \right),a\left( {{\theta _1}} \right), \ldots a\left( {{\theta _{L - 1}}} \right)} \right]$ is an $M \times L$ array matrix, signal vector and noise vector are independent.
$s\left( k \right) = \mathop {\left[ {\mathop s\nolimits_0 \left( k \right),\mathop s\nolimits_1 \left( k \right), \ldots ,\mathop s\nolimits_{L-1} \left( k \right)} \right]}\nolimits^T $ is an incident signal vector with zero mean value, $n\left( k \right)$ is the Gaussian noise vector with zero mean value and $\sigma ^2  I_N $ covariance matrix.
Then, the $M \times M$ spatial covariance matrix of the received signal vector in Eq. (9) can be obtained as 
\begin{equation}
R = E\left\{ {y\left( k \right)y^H \left( k \right)} \right\}
\end{equation}
As the spatial covariance matrix is Hermitian, the eigen-decomposition form can be represented as
\begin{equation}
R{\rm{ =  }}{U_s}{\Lambda _s}{\rm{ }}U_s^H + {\rm{ }}{U_n}{\Lambda _n}{\rm{ }}U_n^H 
\end{equation}
where ${U_s}=\left [ u_1, u_2,\dots ,u_L\right ]$ is an $M \times L$ matrix representing signal subspace, ${U_n}=\left [ u_{L+1}, u_{L+2},\dots ,u_M\right ] $ is a group of the orthogonal base vectors of noise subspace.

The object of the direction angle estimation is
\begin{equation}
\begin{array}{c}
{\theta _{MUSIC}} = \mathop {\arg \min }\limits_\theta  {\rm{ }}{a^H}\left( \theta  \right){\rm{ }}{U_n}{\rm{ }}U_n^Ha\left( \theta  \right)
\end{array}
\end{equation}
Analogously, direction angle estimation can also be represented in terms of its reciprocal to obtain peaks; that is,
\begin{equation}
\begin{array}{c}
\mathop P\nolimits_{MUSIC}  = \frac{1}{{\mathop a\nolimits^H \left( \theta  \right)\mathop U\nolimits_n \mathop U\nolimits_n^H a\left( \theta  \right)}}
\end{array}
\end{equation}
%By searching in direction parameter space, $K$ estimation values which make (5) be peak will be obtained.
In conventional MUSIC algorithm, the distribution of received signal vector ${y\left( k \right)}$ is unknown. Therefore, the spatial covariance matrix in Eq. (10) can be estimated by the sample average, that is $R \approx \frac{1}{N}\sum_{k = 1}^N {y\left( k \right) y^H \left( k \right)} $ where $N$ indicates the number of samples.

\section{QUANTUM ALGORITHM FOR DOA ESTIMATION}
\label{sec_mathematical description}
In this section, we propose a quantum algorithm for MUSIC-based DOA estimation. We first characterize the quantum subroutine for the reconstruction of the spatial covariance matrix in Sec.\ref{sec_mathematical description} A. Next, in Sec.\ref{sec_mathematical description} B, the variational quantum density matrix eigensolver (VQDME) algorithm is designed for the eigen-decomposition in MUSIC algorithm. In this subroutine, we design a novel cost function, which can be effectively computed in a quantum computer. Then, we present a quantum labeling operation based on a set of quantum states for finding directions satisfying Eq. (12) or (13). Finally, the combination of these parts can implement the quantum DOA estimation in hybrid massive MIMO systems.

%\subsection{A quantum subroutine for $\mathop {\hat R}\nolimits_{xx} $}
\subsection{The quantum subroutine for the reconstruction of the spatial covariance matrix}

In this subroutine, to obtain the vector form of the spatial covariance matrix in Eq. (8), we assume that $P$ is stored in quantum random access memory \cite{24} with a suitable data structure of binary trees. Then, the quantum state $\left| P \right\rangle $ can be prepared as 
\begin{equation}
U_P :\left| 0 \right\rangle  \to \frac{1}{{{\left\| P \right\|}_2 }}\sum\nolimits_{i = 0}^{Q - 1} {P_i } \left| i \right\rangle   
\end{equation}
Next, we design an efficient quantum circuit to prepare any row of $A$. As every row of $A$ can be presented as Kronecker product of the steering vector and its conjugate $a_q  = a\left( {\theta^{\left( q \right)} } \right)\otimes a^* \left( {\theta^{\left( q \right)} } \right),q=1,2,\dots ,Q$. Therefore, we first design a quantum circuit to prepare $\left | a\left ( \theta ^{q}  \right )   \right \rangle$, the detailed circuit can be shown in Fig. 3.
\begin{figure*}
\includegraphics[width=4.7in]{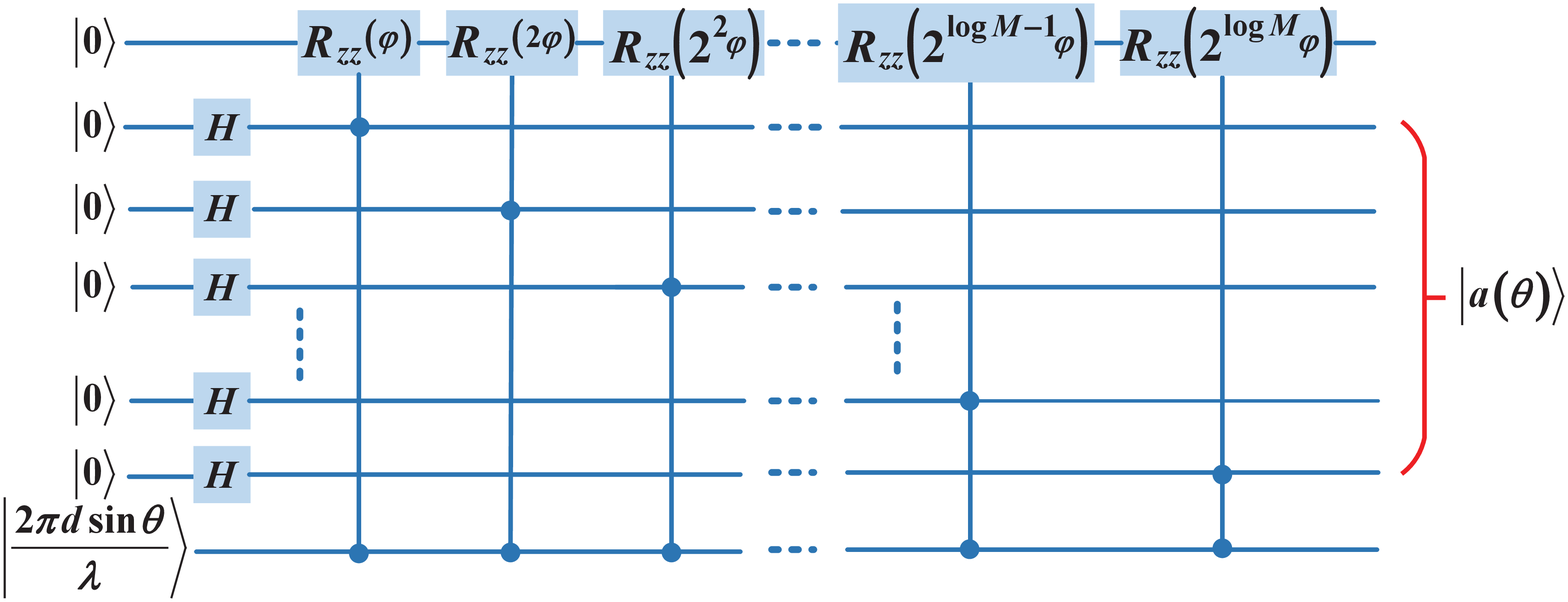}
\caption{Quantum circuit for steering vector and $\mathop R\nolimits_{ZZ} \left( {\mathop \varphi \nolimits_i } \right) = \left( {\begin{array}{*{20}{c}}
{\mathop e\nolimits^{j\mathop \varphi \nolimits_i } }&0\\
0&{\mathop e\nolimits^{j\mathop \varphi \nolimits_i } }
\end{array}} \right)$}
\label{steer}
\end{figure*}
Here, we shown an example of preparing a $4 \times 1$ steering vector to demonstrate the quantum circuit implementation.
\begin{widetext}
\begin{eqnarray}
\begin{array}{l}
\frac{1}{2}\left| {\frac{{2\pi d\sin \theta }}{\lambda }} \right\rangle \left( {\left| 0 \right\rangle  + \left| 1 \right\rangle } \right)\left( {\left| 0 \right\rangle  + \left| 1 \right\rangle } \right)\left| 0 \right\rangle  = \frac{1}{2}\left| {\frac{{2\pi d\sin \theta }}{\lambda }} \right\rangle \left( {\left| 0 \right\rangle \left| 0 \right\rangle  + \left| 0 \right\rangle \left| 1 \right\rangle  + \left| 1 \right\rangle \left| 0 \right\rangle  + \left| 1 \right\rangle \left| 1 \right\rangle } \right)\left| 0 \right\rangle \\
 = \frac{1}{2}\left| {\frac{{2\pi d\sin \theta }}{\lambda }} \right\rangle \left( {\left| 0 \right\rangle \left| 0 \right\rangle \left| 0 \right\rangle  + \left| 0 \right\rangle \left| 1 \right\rangle \mathop e\nolimits^{ - j\frac{{2\pi d\sin \theta }}{\lambda }} \left| 0 \right\rangle  + \left| 1 \right\rangle \left| 0 \right\rangle \mathop e\nolimits^{ - j\frac{{2\pi 2d\sin \theta }}{\lambda }} \left| 0 \right\rangle  + \left| 1 \right\rangle \left| 1 \right\rangle \mathop e\nolimits^{ - j\frac{{2\pi 3d\sin \theta }}{\lambda }} \left| 0 \right\rangle } \right) \\
=\frac{1}{2}\left| {\frac{{2\pi d\sin \theta }}{\lambda }} \right\rangle \left( {\left| 0 \right\rangle \left| 0 \right\rangle  + \mathop e\nolimits^{ - j\frac{{2\pi d\sin \theta }}{\lambda }} \left| 0 \right\rangle \left| 1 \right\rangle  + \mathop e\nolimits^{ - j\frac{{2\pi 2d\sin \theta }}{\lambda }} \left| 1 \right\rangle \left| 0 \right\rangle  + \mathop e\nolimits^{ - j\frac{{2\pi 3d\sin \theta }}{\lambda }} \left| 1 \right\rangle \left| 1 \right\rangle } \right)\left| 0 \right\rangle \\
 = \left| {\frac{{2\pi d\sin \theta }}{\lambda }} \right\rangle \left| {a\left( \theta  \right)} \right\rangle \left| 0 \right\rangle 
\end{array}
\end{eqnarray}
\end{widetext}
Thus, any row of $A$ can be prepared in following quantum circuit Fig. 4, and then we define two unitary mappings $\mathop U\nolimits_{\rm M} $ and $\mathop U\nolimits_{\rm N} $ as follows
\begin{figure*}
\includegraphics[width=5in]{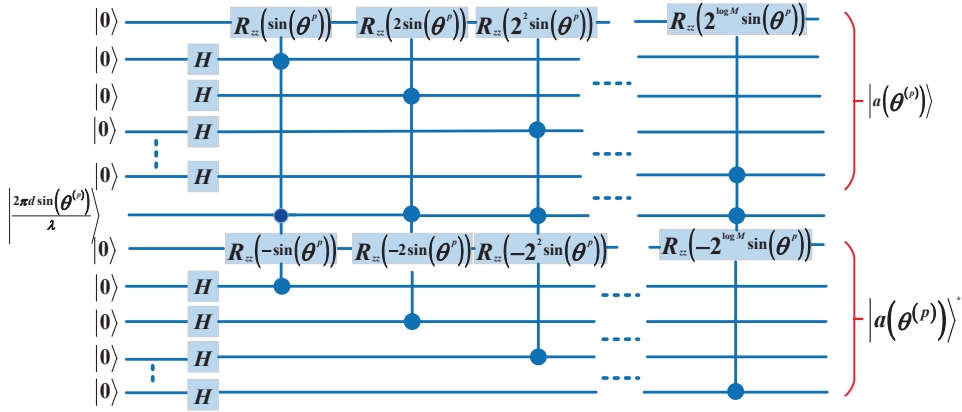}
\caption{Quantum circuit implementing any row of $A$}
\label{fig_chain5566}
\end{figure*}
\begin{equation}
\begin{array}{l}
U_{\rm M} :\left| i \right\rangle \left| 0 \right\rangle  \to \frac{1}{{{\left\| {A_i } \right\|}_2 }}\sum\nolimits_{j = 0}^{M^2  - 1} {A_{ij} \left| i \right\rangle \left| j \right\rangle } =\sum\nolimits_{j = 0}^{M^2  - 1} {\frac{A_{ij}}{M}  \left| i \right\rangle \left| j \right\rangle } 
\end{array} 
\end{equation}
\begin{equation}
\begin{array}{l}
\mathop U\nolimits_{\rm N} :\left| 0 \right\rangle \left| j \right\rangle  \to \frac{1}{{\mathop {\left\| A \right\|}\nolimits_F }}\sum\nolimits_{i = 0}^{Q - 1} {\mathop {\left\| {\mathop A\nolimits_i } \right\|}\nolimits_2 \left| i \right\rangle \left| j \right\rangle } \\
 = \frac{1}{{M\sqrt Q }}\sum\nolimits_{i = 0}^{Q - 1} {M\left| i \right\rangle \left| j \right\rangle } 
\end{array}
\end{equation}
where ${A_i }$ is the $i$th row of $A$. Obviously, $U_{\rm N} $ can be implemented with the complexity ${\rm O}\left( \log Q \right)$ , and based on circuit Fig. 3 and 4, the implementation of $U_{\rm M} $ can be performed in the complexity ${\rm O}\left(\mathrm{ poly } \log MQ \right)$. Moreover, we can obtain the following normalization form of $A$,
\begin{equation}
\left\langle i \right|\left\langle 0 \right|\mathop U\nolimits_{\rm M}^H \mathop U\nolimits_{\rm N} \left| 0 \right\rangle \left| j \right\rangle  = \frac{{\mathop A\nolimits_{ij} }}{{\mathop {\left\| A \right\|}\nolimits_F }} = \frac{{\mathop A\nolimits_{ij} }}{{M\sqrt Q }}
\end{equation}
Then, the quantum state form of ${\hat r}$ can be represented
\begin{widetext}
\begin{eqnarray}
\left| {\hat r} \right\rangle  \propto {\left( {A^H A + \sigma^2 I_{M^2 } } \right)}^{ - 1} A^H \left| P \right\rangle  = 
\sum\nolimits_{i = 1}^\gamma  {\frac{{{\alpha_i \sigma }_i }}{{\sigma _i^2  + \sigma ^2 }}} \left| {v_i } \right\rangle 
\end{eqnarray}
\end{widetext}
where $\alpha_i  = \left\langle {u_i } \right|\left. P \right\rangle$ and $\gamma $ is the rank of $A$, and $\left| {v_i } \right\rangle $, $\left| {u_i } \right\rangle $ and $\sigma_i $ are the right, left singular vectors and corresponding singular value, respectively.
By considering $U_{\rm M} $ and $U_{\rm N} $, we prove the following singular vector transition theorem,
\begin{theorem}
Given a $Q \times M^{2}$ matrix $A$ and the quantum state $ {\textstyle \sum_{i}^{}} \beta _{i} \left | u_i  \right \rangle $, there exists a quantum algorithm implementing the following mapping 
\begin{equation}
\mathrm{T} : {\textstyle \sum_{i}^{}} \beta _{i} \left | u_i  \right \rangle \longrightarrow {\textstyle \sum_{i}^{}} \beta _{i} \left | v_i  \right \rangle
\end{equation}
and the complexity can be approximately estimated as ${\rm O}\left( \frac{\left \|  A\right \| _{F} \mathrm{poly} \left ( \mathrm{log} QM \right ) }{\varepsilon }  \right)$.
\end{theorem}
See {\bfseries{Appendix A}} for details of the singular vector transition theorem.
Therefore, based on the definition of  $U_{\rm M} $, $U_{\rm N} $ and the above theorem, the detailed quantum algorithm is as follows:
\\
(1) Prepare the state as follows
\begin{equation}
\left| {\hat \varphi } \right\rangle {\rm{ = }}\left| 0 \right\rangle \left| P \right\rangle {\rm{ = }}\sum\nolimits_{i = 1}^{Q} {\mathop \alpha \nolimits_i \left| 0 \right\rangle \left| {\mathop u\nolimits_i } \right\rangle }  
\end{equation}
(2) Apply the quantum singular value estimation (QSVE) technique \cite{24}, we have the state
\begin{equation}
\left| {\hat \varphi } \right\rangle {\rm{ = }}\sum\nolimits_{i = 1}^{Q} {\mathop \alpha \nolimits_i \left| {\mathop \sigma \nolimits_i } \right\rangle \left| {\mathop u\nolimits_i } \right\rangle } 
\end{equation}
(3) Add a register with ${\left| 0 \right\rangle }$ and perform the controlled rotation operation, controlled by the register storing the singular value, we have the following state
\begin{equation}
\begin{array}{l}
\left| {\hat \varphi } \right\rangle  = \\
\sum\nolimits_{i = 1}^{Q} {\mathop \alpha \nolimits_i \left( {\frac{{C\mathop \sigma \nolimits_i }}{{\mathop \sigma \nolimits_i^2  + \mathop \sigma \nolimits^2 }}\left| 0 \right\rangle  + \sqrt {1 - \mathop {\left( {\frac{{C\mathop \sigma \nolimits_i }}{{\mathop \sigma \nolimits_i^2  + \mathop \sigma \nolimits^2 }}} \right)}\nolimits^2 } \left| 1 \right\rangle } \right)\left| {\mathop \sigma \nolimits_i } \right\rangle \left| {\mathop u\nolimits_i } \right\rangle } 
\end{array} 
\end{equation}
where $C$ is a constant, that is $C = {\raise0.7ex\hbox{$1$} \!\mathord{\left/
 {\vphantom {1 {\mathop {\max }\limits_i \frac{{\mathop \sigma \nolimits_i }}{{\mathop \sigma \nolimits_i^2  + \mathop \sigma \nolimits^2 }}}}}\right.\kern-\nulldelimiterspace}
\!\lower0.7ex\hbox{${\mathop {\max }\limits_i \frac{{\mathop \sigma \nolimits_i }}{{\mathop \sigma \nolimits_i^2  + \mathop \sigma \nolimits^2 }}}$}}$.
\\
(4) Uncomputing the second register and  measure the first register $\left| 0 \right\rangle $ with the probability $P\left( {\left| 0 \right\rangle } \right)$, we have the state
\begin{equation}
\left| {\hat \varphi } \right\rangle {\rm{ = }}\frac{1}{{\sqrt {\sum\nolimits_{i = 1}^{\gamma } {\mathop {\left| {\mathop \alpha \nolimits_i } \right|}\nolimits^2 \mathop {\left( {\frac{{\mathop \sigma \nolimits_i }}{{\mathop \sigma \nolimits_i^2  + \mathop \sigma \nolimits^2 }}} \right)}\nolimits^2 } } }}\sum\nolimits_{i = 1}^{\gamma }{\frac{{\mathop \alpha \nolimits_i \mathop \sigma \nolimits_i }}{{\mathop \sigma \nolimits_i^2  + \mathop \sigma \nolimits^2 }}} \left| {\mathop u\nolimits_i } \right\rangle  
\end{equation}
where $P\left( {\left| 0 \right\rangle } \right) = \sum\nolimits_{i = 1}^{\gamma } {\mathop {\left| {\mathop \alpha \nolimits_i } \right|}\nolimits^2 \mathop {\left( {\frac{{\mathop {C\sigma }\nolimits_i }}{{\mathop \sigma \nolimits_i^2  + \mathop \sigma \nolimits^2 }}} \right)}\nolimits^2 } $.
\\
(5) Perform the unitary transformation ${\rm T}:\left| {\mathop u\nolimits_i } \right\rangle  \to \left| {\mathop v\nolimits_i } \right\rangle $, we can obtain
\begin{equation}
\left| {\hat r} \right\rangle  = \left| {\hat \varphi } \right\rangle {\rm{ = }}\frac{1}{{\sqrt {\sum\nolimits_{i = 1}^{\gamma } {\mathop {\left| {\mathop \alpha \nolimits_i } \right|}\nolimits^2 \mathop {\left( {\frac{{\mathop \sigma \nolimits_i }}{{\mathop \sigma \nolimits_i^2  + \mathop \sigma \nolimits^2 }}} \right)}\nolimits^2 } } }}\sum\nolimits_{i = 1}^{\gamma } {\frac{{\mathop \alpha \nolimits_i \mathop \sigma \nolimits_i }}{{\mathop \sigma \nolimits_i^2  + \mathop \sigma \nolimits^2 }}} \left| {\mathop v\nolimits_i } \right\rangle  
\end{equation}
where more details of ${\rm T}$ is deferred to {\bfseries{Appendix A}}. Thus, we can obtain the quantum state $\left| {\hat \varphi } \right\rangle $, which is proportional to the vector form of the spatial covariance matrix. Subsequently, for the goal of eigen-decomposition in MUSIC, we transform $\left| {\hat \varphi } \right\rangle $ into a density matrix based on Schmidt decomposition theorem,

\begin{theorem}
Given compound system $\left| \chi  \right\rangle $, there exist subsystems $A$ and $B$ so that 
\begin{equation}
\left| \chi  \right\rangle  = \sum\nolimits_i {\lambda_i } \left| {\varphi_A } \right\rangle \left| {\varphi_B } \right\rangle 
\end{equation}
where $\lambda_i$ is the Schmidt coefficient and $\sum\nolimits_i \lambda_i^2   = 1$, and $\left|\varphi_A \right\rangle$ and $\left| \varphi_B  \right\rangle$ are the standard orthogonal basises of subsystems $A$ and $B$, respectively.
\end{theorem}

By Schmidt decomposition theorem, $\left| {\hat \varphi } \right\rangle $ can be rewritten as follows

\begin{equation}
\begin{array}{c}
\left| {\hat \varphi } \right\rangle {\rm{ = }}\frac{1}{{\sqrt {\sum\nolimits_{i = 1}^{\gamma } {{{\left| {{\alpha _i}} \right|}^2}{{\left( {\frac{{{\sigma _i}}}{{\sigma _i^2 + {\sigma ^2}}}} \right)}^2}} } }}\sum\nolimits_{i = 1}^{\gamma } {\frac{{{\alpha _i}{\sigma _i}}}{{\sigma _i^2 + {\sigma ^2}}}} \left| {{\rm{ }}{v_i}} \right\rangle\\ 
{\rm{ = }}\sum\nolimits_{j = 1}^{\gamma } {\sigma _j^{\hat r}} {{\left| {{\rm{ }}u_j^{\hat r}} \right\rangle }_A}{{\left| {{\rm{ }}v_j^{\hat r}} \right\rangle }_B}
\end{array}
\end{equation}
where $\left|  u_j^{{\hat r} } \right\rangle $ and $\left| v_j^{\hat r}  \right\rangle$ are left and right singular vectors of $\left| {\hat r} \right\rangle $, $\sigma_j^{\hat r}$ is corresponding singular value, respectively. As the spatial covariance matrix is positive definite, its singular vectors are the same as eigenvectors, namely, $\left| u_j^{\hat r} \right\rangle_A = \left| v_j^{\hat r} \right\rangle _B$. Therefore, we take the partial trace on the $B$ register of $\left| {\hat \varphi } \right\rangle \left\langle {\hat \varphi } \right|$ and the density matrix representation $\mathop \rho \nolimits_{\hat R} $ can be denoted as
\begin{equation}
\rho_{\hat R}  = {{\rm tr} }_2 \left( {\left| {\hat \varphi } \right\rangle \left\langle {\hat \varphi } \right|} \right) = \sum\nolimits_{i = 1}^{\gamma }  {{\left( {\sigma_j^{\hat r} } \right)}^2 } {\left| {u_j^{\hat r} } \right\rangle }_A \left\langle {u_j^{\hat r} } \right|
\end{equation}
where $\rho_{\hat R}$ have the same eigenvectors as the spatial covariance matrix.

\subsection {The variational quantum density matrix eigensolver for MUSIC}
Here, we propose the variational quantum eigensolver for the density matrix, which variationally learns the largest nozero eigenvalues of the density matrix as well as a gate sequence $V\left( \theta  \right)$ that
prepares the corresponding eigenvectors. Before elaborately introuducing our variational quantum eigensolver, we first briefly review following von Neumann theorem,

\begin{theorem}
Given a symmetric matrix $H\in R^{M\times M} $, for $k \in \left\{ {1,2, \ldots ,M} \right\}$, there exists the following form
\begin{equation}
\sum\nolimits_{i = 1}^k {\mathop \lambda \nolimits_i }  = \mathop {\max }\limits_{\mathop V\nolimits^H V = I} {\mathop{\rm tr}\nolimits} \left( {\mathop V\nolimits^H HV} \right) = \mathop {\max }\limits_{\left\langle {\mathop v\nolimits_i } \right|\left. {\mathop v\nolimits_j } \right\rangle  = \mathop \delta \nolimits_{ij} } \left\langle {\mathop v\nolimits_i } \right|H\left| {\mathop v\nolimits_j } \right\rangle 
\end{equation}
where $\lambda _{1},\lambda _{2},\dots \lambda _{k}$ are eigenvalues of $H$ and $V = \left[ {\mathop v\nolimits_1 ,\mathop v\nolimits_2 , \ldots \mathop v\nolimits_k } \right]$ are corresponding eigenvectors.
\end{theorem}

Based on von Neumann theorem and Subspace-search variational quantum eigensolver \cite{51}, our variational quantum eigensolver can be depicted in {\bfseries{Algorithm 1}}
\begin{algorithm}[H]
	\caption{\bf{: The variational quantum density matrix eigensolver}}{{\bfseries{Input :}} the density matrix $\rho_{\hat R}$, a set of positive real weights $\mathop q\nolimits_1  > \mathop q\nolimits_2  >  \cdots  > \mathop q\nolimits_L  > 0$}\\
\emph{step 1:} Construct an ansatz circuit $V\left ( \overrightarrow{\theta }  \right )$ and choose input states $\mathop {\left\{ {\left| {\mathop \varphi \nolimits_i } \right\rangle } \right\}}\nolimits_{i = 1}^L $, which are mutually orthogonal quantum states.\\
\emph{step 2:} Define an objective function $L\left( \overrightarrow{\theta}  \right) = \left ( {\textstyle \sum_{i=1}^{L}} q_i   \right ) \sum\nolimits_{i = 1}^L\frac{{q_i }}{\left ( {\textstyle \sum_{i=1}^{L}} q_i   \right ) }  \left\langle {\varphi_i } \right|V^H \left( \overrightarrow{\theta}   \right)\rho_{\hat R} V\left( \overrightarrow{\theta}   \right)\left| {\varphi_i } \right\rangle   $.\\
\emph{step 3:} Rewritten the objective function $L\left ( \overrightarrow{\theta }  \right )$ as $\left ( {\textstyle \sum_{i=1}^{L}} q_i   \right )C\left ( \overrightarrow{\theta }  \right )$.\\ %where $C\left ( \overrightarrow{\theta }  \right ) =\mathrm{tr} \left ( V^{H}\left ( \overrightarrow{\theta }  \right ) \rho _{\hat{R} } V\left ( \overrightarrow{\theta }  \right )  \rho _{f}  \right ) $\\
\emph{step 4:} Compute $C\left ( \overrightarrow{\theta }  \right )$ in a quantum computer and apply the classical optimization algorithm to maximize $C\left ( \overrightarrow{\theta }  \right )$. The optimal $\overrightarrow{\theta}$ can be represented as $\overrightarrow{\theta }^{\ast } $.\\
{\bfseries{Output :}} $\left \{  V\left ( \overrightarrow{\theta }^{\ast }   \right ) \left | \varphi _{i}   \right \rangle \right \} _{i=1}^{L} $ are approximate eigenvectors.
\end{algorithm}

In step 1, given a series of parameter vectors $\overrightarrow{\theta } =\left ( \theta _{1}  ,\theta _{2},\dots \theta _{m}  \right ) $, the quantum circuit $V$ is defined as 
\begin{equation}
V\left ( \overrightarrow{\theta }  \right ) =V_{m} \left ( \theta _{m}  \right )V_{m-1} \left ( \theta _{m-1}  \right )\dots V_{1} \left ( \theta _{1}  \right ) 
\end{equation}
where the ansatz circuit can be shown in Fig. 5. Note that the number of parameters is logarithmically proportional to the dimension of the density matrix. These parameterized quantum circuits have shown significant potential power in quantum neural network and quantum circuit Born machines \cite{52,53,54}.

\begin{figure}
\includegraphics[width=3.5in]{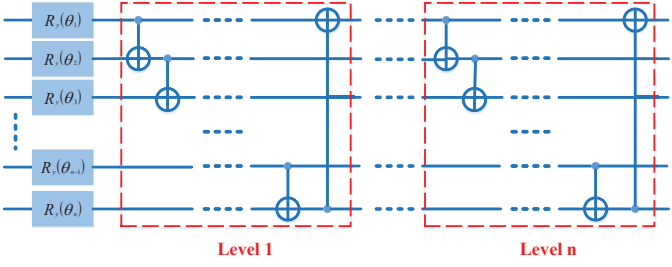}
\caption{The structure of the ansatz circuit.}
\label{fig_chain7788}
\end{figure}

In step 3, $C\left ( \overrightarrow{\theta }  \right )$ can be denoted as
\begin{equation}
C\left ( \overrightarrow{\theta }  \right ) = \sum\nolimits_{i = 1}^L\frac{{q_i }}{\left ( {\textstyle \sum_{i=1}^{L}} q_i   \right ) }  \left\langle {\varphi_i } \right|V^H \left( \overrightarrow{\theta}   \right)\rho_{\hat R} V\left( \overrightarrow{\theta}   \right)\left| {\varphi_i } \right\rangle   
\end{equation}
Further, we can transform $C\left ( \overrightarrow{\theta }  \right )$ into a trace form as follows
\begin{equation}
C\left ( \overrightarrow{\theta }  \right ) =\mathrm{tr} \left ( V^{H}\left ( \overrightarrow{\theta }  \right ) \rho _{\hat{R} } V\left ( \overrightarrow{\theta }  \right )  \rho _{f}  \right ) 
\end{equation}
where $\rho _{f} = {\textstyle \sum_{i=1}^{L}} \frac{q_i}{{\textstyle \sum_{i=1}^{L}}q_i} \left | \varphi _{i}   \right \rangle\left \langle \varphi _{i}  \right |  $. Unlike existing variational quantum eigensolvers, the density matrix $\rho_{\hat R}$ can not be represented as a linear combination of unitary matrices. This is because the density matrix is the output of the quantum algorithm in the previous quantum subroutine, we do not have any prior information on $\rho_{\hat R}$ without quantum tomography. However, we transform the expection form in the objective function into the trace of two density matrices. Thus, in our quantum algorithm, a linear combination of unitary matrices is not required and the objective function $C\left ( \overrightarrow{\theta }  \right )$ can effectively be implemented in Destructive Swap Test, which is shown in Fig. 6. Moreover, the detailed quantum implementation scheme is depicted in Fig. 7, and for the implementation, we consider the following a $4 \times 4$ density matrix (using two qubit).

\emph{Example:} An random density matrix $\rho_{in}$ with four fixed eigenvalues $\lambda _{1} = 0.4,\lambda _{2} = 0.3,\lambda _{3} = 0.2,\lambda _{4} = 0.1$, a group of positive real weights are $q _{1} = 4,q _{2} = 3,q _{3} = 2,q _{4} = 1$ and $\left | \varphi _{1}   \right \rangle =\left | 1  \right \rangle ,\left | \varphi _{2}   \right \rangle =\left | 2  \right \rangle,\left | \varphi _{3}   \right \rangle =\left | 3  \right \rangle,\left | \varphi _{4}   \right \rangle =\left | 4  \right \rangle   $. The experiment's results of the VQDME implementation are shown in Fig. 8. As shown in the above figures, we can obtain four eigenvalues and eigenvectors within 200 quantum-classical hybrid iterations, which are very close to the exact ones.

\begin{figure}
\includegraphics[width=3.0in]{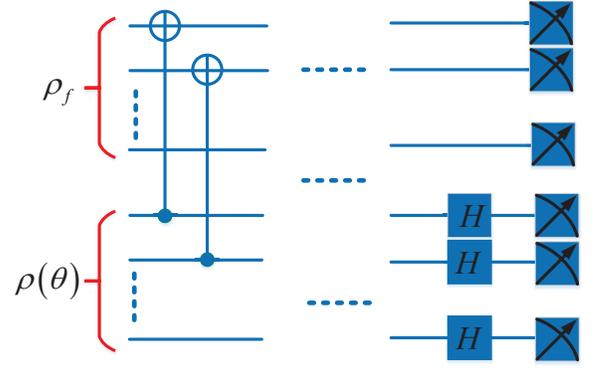}
\caption{Destructive Swap Test for loss function $C$.}
\label{fig_chain111}
\end{figure}

\begin{figure*}
\includegraphics[width=4.5in]{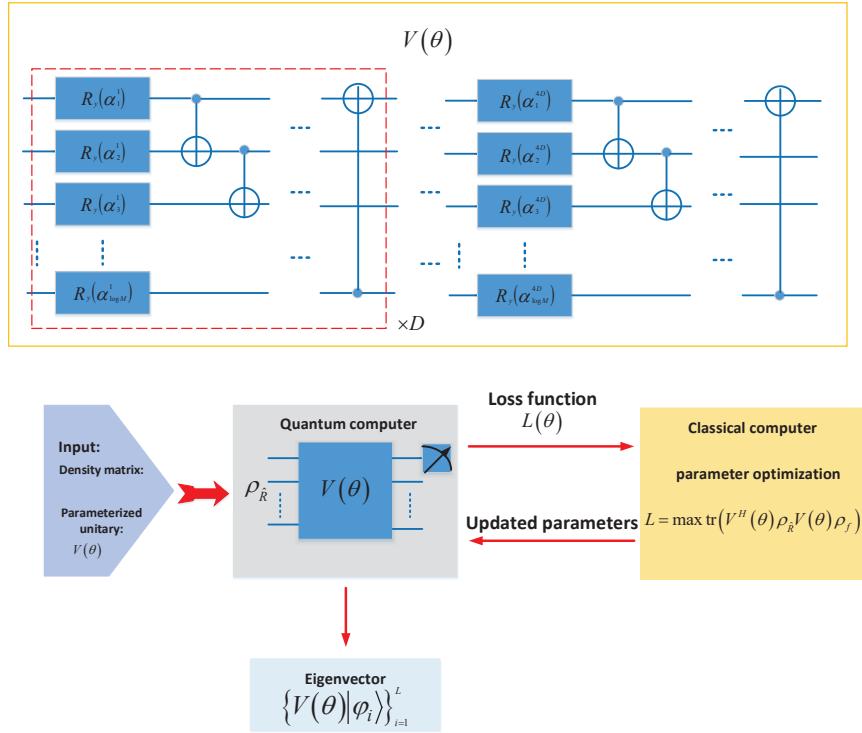}
\caption{Schematic diagram for VQDME algorithm}
\label{fig_chain2222}
\end{figure*}

\begin{figure*}[!htp]
\centering

\subfigure[The convergence to the first eigenvalue.]{
\centering
\includegraphics[width=4.4in]{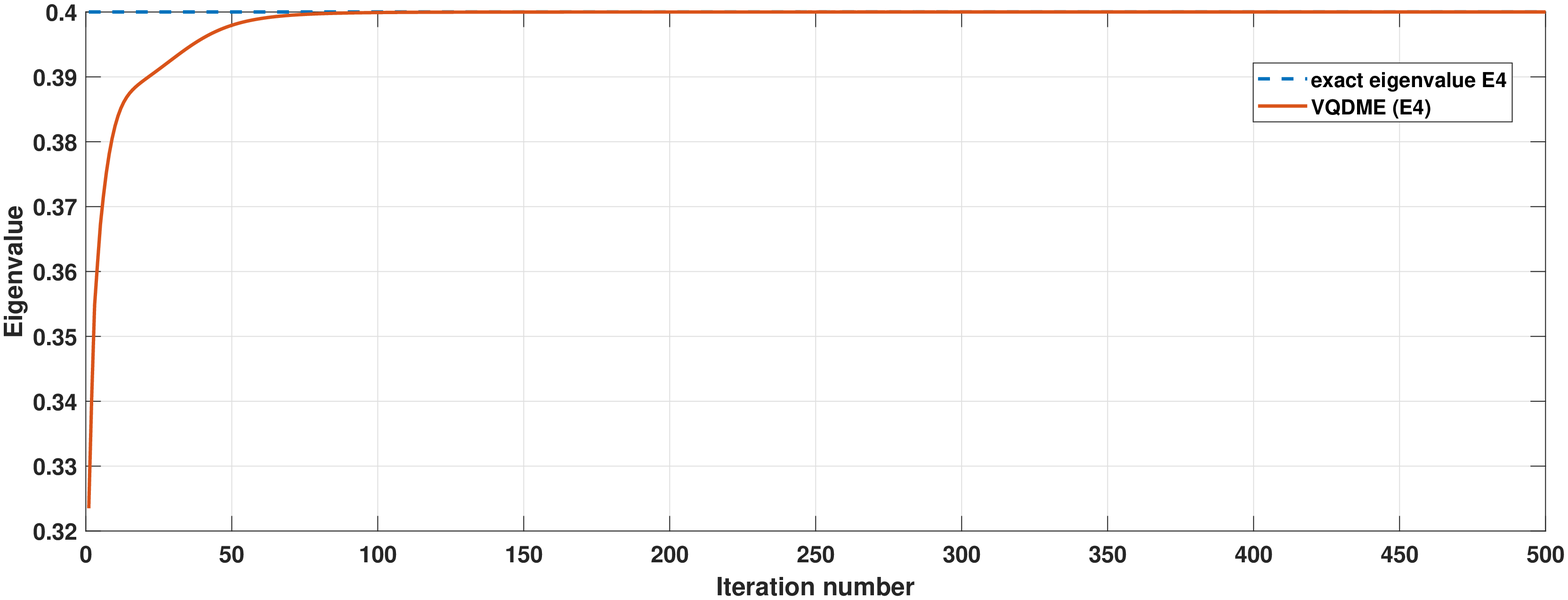}
}
\subfigure[The convergence to the second eigenvalue.]{
\centering
\includegraphics[width=4.4in]{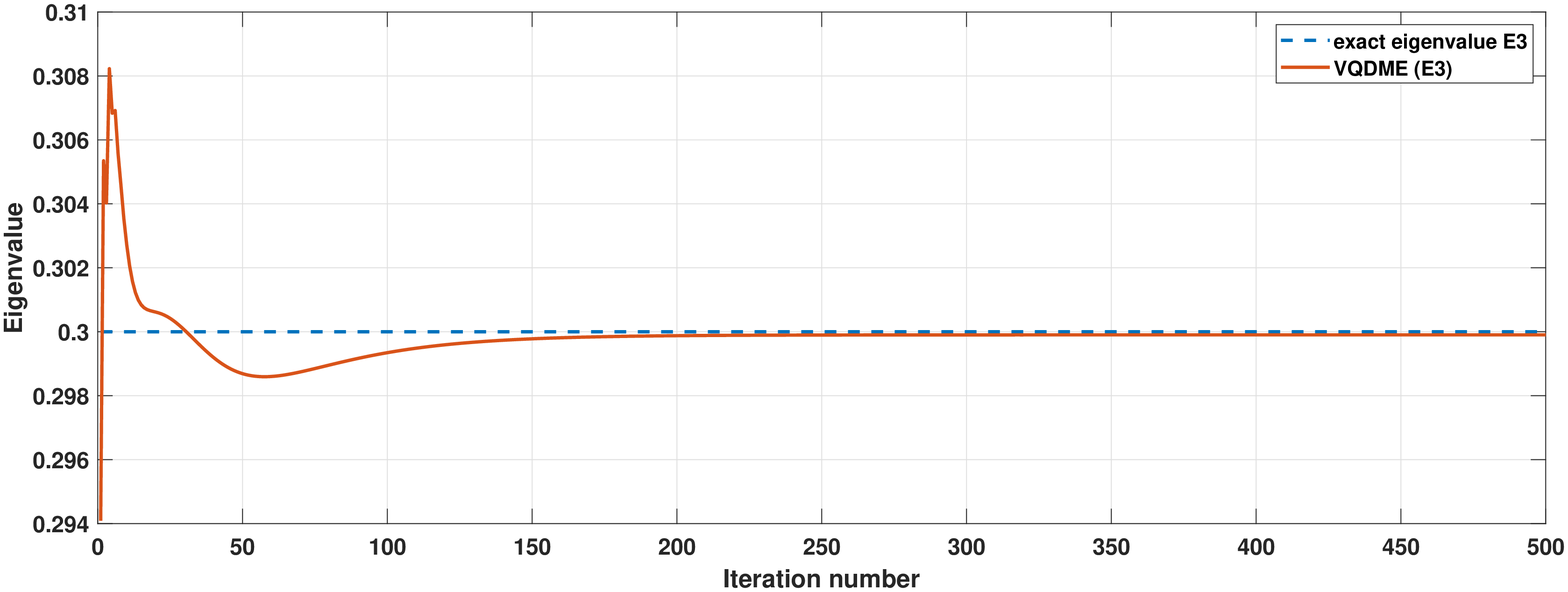}
}%

\subfigure[The convergence to the third eigenvalue.]{
\centering
\includegraphics[width=4.2in]{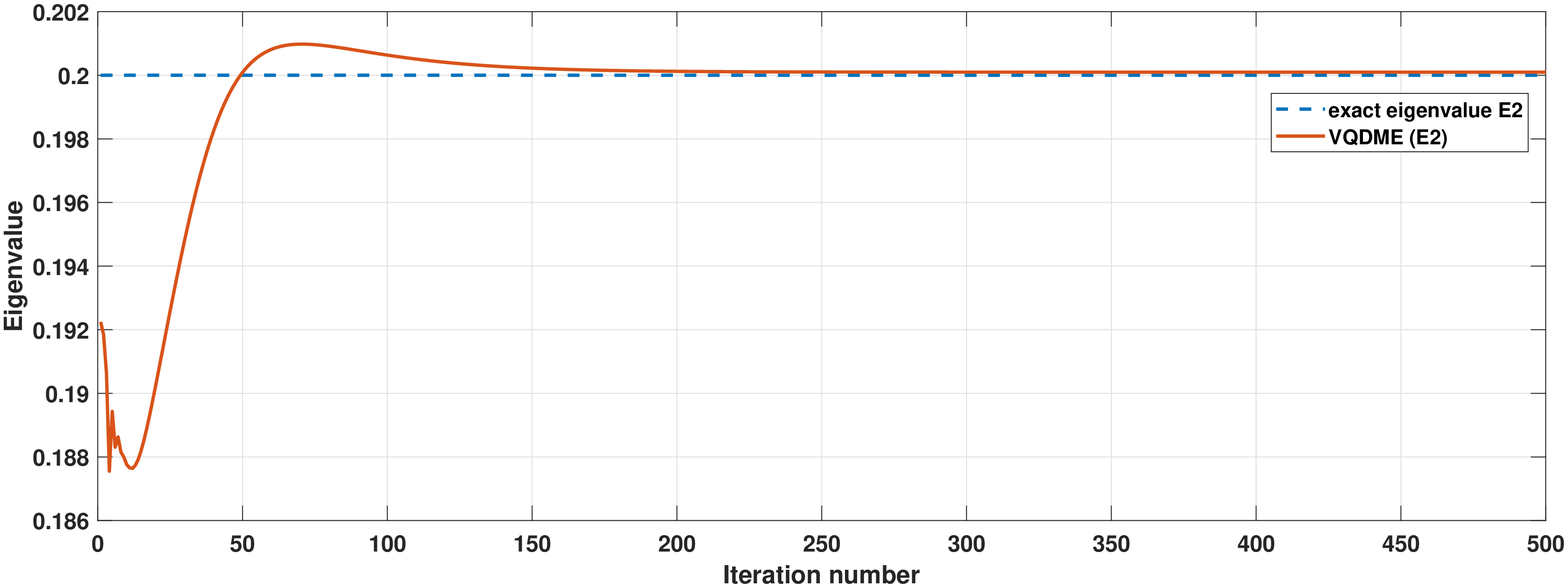}
}%

\subfigure[The convergence to the fourth eigenvalue.]{
\centering
\includegraphics[width=4.2in]{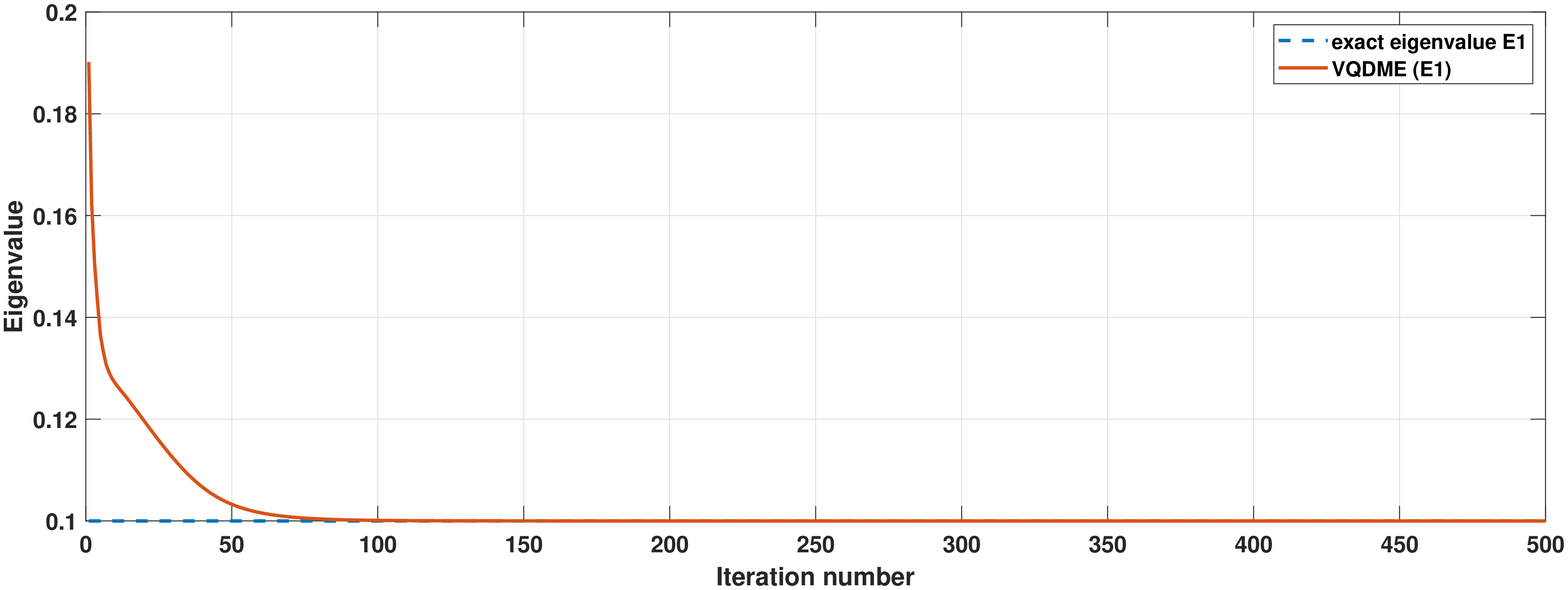}
}%
\centering
\caption{Convergence curve of the VQDME implementation for example}
\end{figure*}

\subsection {The quantum labeling operation for directions searching}

To implement the direction angles estimation, we can rewrite the object (12) as
\begin{equation}
\begin{array}{l}
\mathop \theta \nolimits_{MUSIC}  = \mathop {\min }\limits_\theta  \mathop a\nolimits^H \left( \theta  \right)\mathop U\nolimits_n \mathop U\nolimits_n^H a\left( \theta  \right) = \\
\mathop {\min }\limits_\theta  \mathop a\nolimits^H \left( \theta  \right)\left( {I - \mathop U\nolimits_s \mathop U\nolimits_s^H } \right)a\left( \theta  \right) = \\
\mathop {\min }\limits_\theta  \left( {M - \mathop a\nolimits^H \left( \theta  \right)\mathop U\nolimits_s \mathop U\nolimits_s^H a\left( \theta  \right)} \right) = \\
\mathop {\max }\limits_\theta  \mathop a\nolimits^H \left( \theta  \right)\mathop U\nolimits_s \mathop U\nolimits_s^H a\left( \theta  \right)= \mathop {\max }\limits_\theta{\textstyle \sum_{i=1}^{L}} \left ( \alpha _{i}^{\theta } \right )^2  
\end{array}
\end{equation}
where $\alpha _{i}^{\theta }=a^{H} \left ( \theta  \right ) u_{i} , i=1,2,\dots ,L$ are the projections on the signal subspace.
Although we have implemented the optimal quantum network $V\left ( \overrightarrow{\theta }^{\ast }   \right ) $, it is hard to construct the controlled $V\left ( \overrightarrow{\theta }^{\ast }   \right ) $. Namely, we can not prepare a coherent state of $\left \{ \left | u_{i}  \right \rangle \right \} _{i=1}^{L} $ and the projections on the signal subspace can also be calculated by the quantum parallelism. Fortunately, the projections on entire space can be obtained by the quantum computing, and a quantum labeling algorithm is required to extract the projections on the signal subspace from ones on entire space. Here, we design a quantum labeling operation based on the set $\left \{ \left | \varphi _{i}   \right \rangle  \right \} _{i=1}^{L} $ and the detailed implementation is as follows:
\\
(1) Prepare the coherent quantum state $\left| {\mathop \phi \nolimits_S } \right\rangle $ for the searching space vector set $\left| {a\left( {\mathop \theta \nolimits_n } \right)} \right\rangle $, $n = 1,2, \ldots ,K$, as
\begin{equation}
\begin{array}{l}
\left| {\mathop \phi \nolimits_S } \right\rangle  = \frac{1}{{\sqrt K }}\sum\nolimits_{n = 1}^K {\left| n \right\rangle \left| {a\left( {\mathop \theta \nolimits_n } \right)} \right\rangle }  = \\
\frac{1}{{\sqrt K }}\sum\nolimits_{n = 1}^K {\left| n \right\rangle \sum\nolimits_{i = 1}^M {\mathop \alpha \nolimits_i^n \left| {\mathop u\nolimits_i^{\hat r} } \right\rangle } } 
\end{array}  
\end{equation}
where ${\left| {\mathop u\nolimits_i^{\hat r} } \right\rangle }$ is the eigenvector of $\mathop \rho \nolimits_{\hat R} $ and $\mathop \alpha \nolimits_i^n  = \left\langle {\mathop u\nolimits_i^{\hat r} } \right.\left| {a\left( {\mathop \theta \nolimits_n } \right)} \right\rangle ,i=1,2,\dots ,M$ is the projections on the entire space of searching space vector.
\\\\
(2) Perform the inverse optimal quantum network $\mathop V\nolimits^H \left( {\mathop \theta \nolimits^* } \right)$ on the second register, we have the following state
\begin{equation}
\left| {\mathop \phi \nolimits_S } \right\rangle  = \frac{1}{{\sqrt K }}\sum\nolimits_{n = 1}^K {\left| n \right\rangle \sum\nolimits_{i = 1}^M {\mathop \alpha \nolimits_i^n \left| {\mathop \varphi \nolimits_i } \right\rangle } } 
\end{equation}
(3) Append a single qubit label register with the initial $\left| 0 \right\rangle $. As we select some certain orthogonal basis vectors as ${\left| {\mathop \varphi \nolimits_i } \right\rangle }$, without loss of generality let $ \left \{ \left| {\mathop \varphi \nolimits_i } \right\rangle \right \} _{i=1}^{L}$ be the computational basis set $\left \{ \left| i \right\rangle  \right \}  _{i=1}^{L} $, then we can construct the labeling mapping as
\begin{equation}
 {\textstyle \sum_{k=1}^{L}} \left | k  \right \rangle \left \langle k \right | \otimes X+\left ( I_{M}- {\textstyle \sum_{k=1}^{L}} \left | k  \right \rangle \left \langle k \right | \right )\otimes I_2 
\end{equation}
and the operation on $ \left \{ \left| {\mathop \varphi \nolimits_i } \right\rangle \right \} _{i=1}^{L}$ can be denoted as
\begin{equation}
CU:\left| {\mathop \varphi \nolimits_i } \right\rangle \mathop {\left| 0 \right\rangle }\nolimits^{{\mathop{\rm label}\nolimits} }  \to \left| {\mathop \varphi \nolimits_i } \right\rangle \mathop {\left| 1 \right\rangle }\nolimits^{{\mathop{\rm label}\nolimits} } ,{\kern 1pt} {\kern 1pt} {\kern 1pt} {\kern 1pt} {\kern 1pt} {\kern 1pt} i = 1,2, \ldots ,L
\end{equation}
\begin{equation}
CU:\left| {\mathop \varphi \nolimits_i } \right\rangle \mathop {\left| 0 \right\rangle }\nolimits^{{\mathop{\rm label}\nolimits} }  \to \left| {\mathop \varphi \nolimits_i } \right\rangle \mathop {\left| 0 \right\rangle }\nolimits^{{\mathop{\rm label}\nolimits} } ,{\kern 1pt} {\kern 1pt} {\kern 1pt} {\kern 1pt} {\kern 1pt} {\kern 1pt} i = L + 1, \ldots ,M
\end{equation}
Thus, we have the state as follows
\begin{equation}
\begin{array}{l}
CU:\left| {\mathop \phi \nolimits_S } \right\rangle  \to \\
\sum\nolimits_{n = 1}^K {\frac{{\left| n \right\rangle }}{{\sqrt K }}\left( {\sum\nolimits_{i = 1}^L {\mathop \alpha \nolimits_i^n \left| {\mathop \varphi \nolimits_i } \right\rangle \mathop {\left| 1 \right\rangle }\nolimits^{{\mathop{\rm label}\nolimits} }  + \sum\nolimits_{i = L + 1}^M {\mathop \alpha \nolimits_i^n \left| {\mathop \varphi \nolimits_i } \right\rangle \mathop {\left| 0 \right\rangle }\nolimits^{{\mathop{\rm label}\nolimits} } } } } \right)} 
\end{array}
\end{equation}
Therefore, the projections of $\left| {a\left( {\mathop \theta \nolimits_n } \right)} \right\rangle $ on the signal subspace labeled ${\mathop {\left| 1 \right\rangle }\nolimits^{{\mathop{\rm label}\nolimits} } }$ are separated from projections on the noise subspace with label ${\mathop {\left| 0 \right\rangle }\nolimits^{{\mathop{\rm label}\nolimits} } }$.
\\
(4) Measure the label register in ${\mathop {\left| 1 \right\rangle }\nolimits^{{\mathop{\rm label}\nolimits} } }$ with the probability $\mathop P\nolimits_S \left( {\left| 1 \right\rangle } \right)$, then we have the state
\begin{equation}
\left| {\mathop \phi \nolimits_S } \right\rangle  = \frac{1}{{\sqrt {\sum\nolimits_{n = 1}^K {\sum\nolimits_{i = 1}^L {\mathop {\left( {\mathop \alpha \nolimits_i^n } \right)}\nolimits^2 } } } }}\sum\nolimits_{n = 1}^K {\left| n \right\rangle \sum\nolimits_{i = 1}^L {\mathop \alpha \nolimits_i^n \left| {\mathop \varphi \nolimits_i } \right\rangle } } 
\end{equation}
(5) Perform a small amount of sampling on the first register of $\left| {\mathop \phi \nolimits_S } \right\rangle $ with the probability $P\left( {\left| n \right\rangle } \right) = \frac{{\sum\nolimits_{i = 1}^L {\mathop {\left( {\mathop \alpha \nolimits_i^n } \right)}\nolimits^2 } }}{{\sqrt {\sum\nolimits_{n = 1}^K {\sum\nolimits_{i = 1}^L {\mathop {\left( {\mathop \alpha \nolimits_i^n } \right)}\nolimits^2 } } } }}$, and satisfy 
\begin{equation}
\frac{{\sum\nolimits_{i = 1}^L {\mathop {\left( {\mathop \alpha \nolimits_i^n } \right)}\nolimits^2 } }}{{\sqrt {\sum\nolimits_{n = 1}^K {\sum\nolimits_{i = 1}^L {\mathop {\left( {\mathop \alpha \nolimits_i^n } \right)}\nolimits^2 } } } }} \propto \mathop a\nolimits^H \left( \theta  \right)\mathop U\nolimits_s \mathop U\nolimits_s^H a\left( \theta  \right)
\end{equation}
As sample can be more easily selected with higher probability, the most frequently selected samples can represent the direction estimation values that satisfy the object (33).

\subsection{Time complexity and error analysis}

\textit{Time complexity for $\left| {\hat \varphi } \right\rangle $:} The state $\left| P \right\rangle $ can be prepared with the complexity ${\rm O}\left( {{\mathop{\rm poly}\nolimits} \left( {\log Q} \right)} \right)$, $\mathop U\nolimits_{\rm N} $ and $\mathop U\nolimits_{\rm M} $ can be implemented in the runtime ${\rm O}\left( {\log Q} \right)$ and ${\rm O}\left( {{\mathop{\rm poly}\nolimits} \left( {\log QM} \right)} \right)$, respectively. Let the error in SVE be $\mathop \varepsilon \nolimits_p \mathop {\left\| A \right\|}\nolimits_F $, where $\mathop \varepsilon \nolimits_p $ is the error in the phase estimation. We define the function $h\left( \sigma  \right)$ as
\begin{equation}
h\left( \lambda  \right) = \frac{\lambda }{{\mathop \lambda \nolimits^2  + \mathop \sigma \nolimits^2 }},{\kern 1pt} {\kern 1pt} {\kern 1pt} {\kern 1pt} \lambda  \in \left[ {\frac{1}{{\mathop \kappa \nolimits_A }},1} \right]
\end{equation}
where ${\mathop \kappa \nolimits_A }$ is the condition number of $A$. Then, we take the derivative of $h\left( \sigma  \right)$ as
\begin{equation}
h'\left( \lambda  \right) = \left\{ {\begin{array}{*{20}{c}}
{\frac{{\mathop \sigma \nolimits^2  - \mathop \lambda \nolimits^2 }}{{\left( {\mathop \lambda \nolimits^2  + \mathop \sigma \nolimits^2 } \right)}} < 0,\lambda  > \sigma {\kern 1pt} {\kern 1pt} {\kern 1pt} }\\
{\frac{{\mathop \sigma \nolimits^2  - \mathop \lambda \nolimits^2 }}{{\left( {\mathop \lambda \nolimits^2  + \mathop \sigma \nolimits^2 } \right)}} \ge 0,0 \le \lambda  \le \sigma {\kern 1pt} {\kern 1pt} }
\end{array}} \right.
\end{equation}
As the parameter ${\mathop \sigma \nolimits^2 }$ is required to be a number closed to 0 in the reconstruction of the spatial covariance matrix, we can deduce $\sigma  \le \frac{1}{{\mathop \kappa \nolimits_A }}$. Therefore, $h\left( \sigma  \right)$ is demonstrated to be monotonic decreasing with $\left[ {\frac{1}{{\mathop \kappa \nolimits_A }},1} \right]$. Thus, we can estimate
\begin{equation}
\frac{1}{{1 + \mathop \sigma \nolimits^2 }} \le \sqrt {\sum\nolimits_{i = 1}^Q {\mathop {\left| {\mathop \alpha \nolimits_i } \right|}\nolimits^2 \mathop {\left( {\frac{{\mathop \sigma \nolimits_i }}{{\mathop \sigma \nolimits_i^2  + \mathop \sigma \nolimits^2 }}} \right)}\nolimits^2 } }  < \mathop \kappa \nolimits_A 
\end{equation}
, $C = \frac{1}{{\mathop \kappa \nolimits_A }} + \mathop \kappa \nolimits_A \mathop \sigma \nolimits^2 $ and

\begin{widetext}
\begin{eqnarray}
P\left( {\left| 0 \right\rangle } \right) = \sum\nolimits_{i = 1}^Q {\mathop {\left| {\mathop \alpha \nolimits_i } \right|}\nolimits^2 \mathop {\left( {\frac{{\mathop {C\sigma }\nolimits_i }}{{\mathop \sigma \nolimits_i^2  + \mathop \sigma \nolimits^2 }}} \right)}\nolimits^2 }  \ge \mathop C\nolimits^2 \mathop {\left( {\frac{1}{{1 + \mathop \sigma \nolimits^2 }}} \right)}\nolimits^2  = \mathop {\left( {\frac{1}{{\mathop \kappa \nolimits_A }} + \mathop \kappa \nolimits_A \mathop \sigma \nolimits^2 } \right)}\nolimits^2 \mathop {\left( {\frac{1}{{1 + \mathop \sigma \nolimits^2 }}} \right)}\nolimits^2  > \frac{1}{4}\mathop {\left( {\frac{1}{{\mathop \kappa \nolimits_A }} + \mathop \kappa \nolimits_A \mathop \sigma \nolimits^2 } \right)}\nolimits^2
\end{eqnarray}
\end{widetext}
Moreover, we can estimate $\left| {h\left( \lambda  \right) - h\left( {\tilde \lambda } \right)} \right|$ as
\begin{equation}
\begin{array}{l}
\left| {h\left( \lambda  \right) - h\left( {\tilde \lambda } \right)} \right| \approx \frac{{\left| {h\left( \lambda  \right) - h\left( {\tilde \lambda } \right)} \right|}}{{\left| {\lambda  - \tilde \lambda } \right|}}\left| {\lambda  - \tilde \lambda } \right| \approx \left| {h'\left( \lambda  \right)} \right|\left| {\lambda  - \tilde \lambda } \right| < \\
\frac{{\mathop \kappa \nolimits_A \mathop \varepsilon \nolimits_p \mathop {\left\| A \right\|}\nolimits_F }}{{1 + \mathop {\mathop \sigma \nolimits^2 \kappa }\nolimits_A^2 }}
\end{array}
\end{equation}
We define the ideal state $\left| {\mathop \varphi \nolimits^* } \right\rangle $, then we have 
\begin{equation}
\begin{array}{l}
\mathop {\left\| {\left| {\hat \varphi } \right\rangle  - \left| {\mathop \varphi \nolimits^* } \right\rangle } \right\|}\nolimits^2  \le \mathop {\left\| {\sum\nolimits_{i = 1}^Q {\frac{{\mathop \sigma \nolimits_i }}{{\mathop \sigma \nolimits_i^2  + \mathop \sigma \nolimits^2 }}\left| {\mathop v\nolimits_i } \right\rangle  - \sum\nolimits_{i = 1}^Q {\frac{{\mathop {\tilde \sigma }\nolimits_i }}{{\mathop {\tilde \sigma }\nolimits_i^2  + \mathop \sigma \nolimits^2 }}\left| {\mathop v\nolimits_i } \right\rangle } } } \right\|}\nolimits^2 \\
 \le \mathop {\left\| {\sum\nolimits_{i = 1}^Q {\left( {\frac{{\mathop \sigma \nolimits_i }}{{\mathop \sigma \nolimits_i^2  + \mathop \sigma \nolimits^2 }} - \frac{{\mathop {\tilde \sigma }\nolimits_i }}{{\mathop {\tilde \sigma }\nolimits_i^2  + \mathop \sigma \nolimits^2 }}} \right)\left| {\mathop v\nolimits_i } \right\rangle } } \right\|}\nolimits^2  \le \mathop {\left( {\frac{{\mathop \kappa \nolimits_A \mathop \varepsilon \nolimits_p \mathop {\left\| A \right\|}\nolimits_F }}{{1 + \mathop {\mathop \sigma \nolimits^2 \kappa }\nolimits_A^2 }}} \right)}\nolimits^2 
\end{array}
\end{equation}
Let the final error be $\varepsilon $, that is $\left\| {\left| {\hat \varphi } \right\rangle  - \left| {\mathop \varphi \nolimits^* } \right\rangle } \right\| < \varepsilon $, we have
\begin{equation}
\mathop \varepsilon \nolimits_p  = \frac{{\varepsilon \left( {1 + \mathop {\mathop \sigma \nolimits^2 \kappa }\nolimits_A^2 } \right)}}{{\mathop {\mathop \kappa \nolimits_A \left\| A \right\|}\nolimits_F }}
\end{equation}
Considering the complexity of the singular vector transition, applying the amplitude amplification technique \cite{55}, the complexity of implementing $\left| {\hat \varphi } \right\rangle $ is estimated as 
\begin{equation}
\begin{array}{l}
{\rm O}\left( {\frac{{\mathop {\mathop \kappa \nolimits_A^2 \left\| A \right\|}\nolimits_F }}{{\varepsilon \left( {1 + \mathop {\mathop \sigma \nolimits^2 \kappa }\nolimits_A^2 } \right)}}{\mathop{\rm poly}\nolimits} \left( {\log MQ} \right)} \right) = \\
{\rm O}\left( {\frac{{\mathop {\mathop \kappa \nolimits_A^2 \left\| A \right\|}\nolimits_F {\mathop{\rm poly}\nolimits} \left( {\log MQ} \right)}}{\varepsilon }} \right)
\end{array}
\end{equation}

\textit{Time complexity for VQDME:} Let the error of VQDME algorithm be $\varepsilon $.
Based on \cite{39}, we can approximately estimate the complexity of VQDME as 
\begin{equation}
{\rm O}\left( {\frac{{\mathop T\nolimits_{VQDME} \mathrm{poly} \left (\log M  \right ) }}{{\mathop \varepsilon \nolimits^2 }}} \right)
\end{equation}
where ${\mathop T\nolimits_{VQDME} }$ is the iteration number of the optimization. Therefore, we can obtain the signal subspace with ${\rm O}\left( {\frac{{\mathop T\nolimits_{VQDME} \mathrm{poly} \left (\log M  \right ) }}{{\mathop \varepsilon \nolimits^2 }}} \right)$ times quantum-classical hybrid operations.

\textit{Time complexity for quantum labeling:} First, the searching space can be prepared with the complexity ${\rm O}\left( {{\rm{poly}}\log KM} \right)$, and then we can estimate 
\begin{equation}
\mathop P\nolimits_S \left( {\left| 1 \right\rangle } \right) = \frac{1}{K}\sum\nolimits_{n = 1}^K {\sum\nolimits_{i = 1}^L {\mathop {\left| {\mathop \alpha \nolimits_i^n } \right|}\nolimits^2 } }  = \Theta \left( 1 \right)
\end{equation}
Moreover, a small amount of sampling are required for the direction estimation satisfying the object (33). 

Therefore, Combining the above three quantum subroutines, our MUSIC-based DOA estimation algorithm can be implemented with the complexity 
\begin{equation}
{\rm O}\left( {\frac{{\mathop {\mathop \kappa \nolimits_A^2 T_{VQDME}\left\| A \right\|}\nolimits_F {\rm{poly}}\log MKQ}}{{\mathop \varepsilon \nolimits^3 }}} \right){\kern 1pt} {\kern 1pt} {\kern 1pt} {\kern 1pt} {\kern 1pt} {\kern 1pt} {\kern 1pt} {\kern 1pt} {\kern 1pt} {\kern 1pt} {\kern 1pt} 
\end{equation}
Without loss of generality, the amplitude of the element in $A$ is less than or equal to $1$ and $Q\gg M$, thus $\left \|  A\right \| _{F} $ is approximate ${\rm O}\left ( Q \right )$. 
\section{Conclusion}
\label{sec_Conclusion}
In the present study, we design the quantum algorithm for MUSIC-based DOA estimation. In classical DOA, the time complexity for the reconstruction of the spatial covariance matrix, the eigenvalue decomposition and direction parameter search are approximate ${\rm O}\left( {{\rm{poly}}MQK} \right)$. The exponentially large number of array and signal is intricate in classical computers. Compared with classical  MUSIC-based DOA estimation, our quantum algorithm can provide an exponential speedup on $M$ and $K$, and a polynomial speedup on $Q$ over classical counterparts when $T_{VQDME}$, $\kappa _{A}  $ and $\frac{1}{\varepsilon} $ can be approximately estimated as ${\rm O}\left ( \mathrm{poly}\left ( \mathrm{log}MKQ  \right )   \right )$. In our algorithm, we first develop the quantum subroutine for the vector form of the spatial covariance matrix. The subroutine based on SVE can be implemented without the extending Hermitian form of the matrix. Second, a variational quantum density matrix eigensolver (VQDME) is proposed for obtaining signal and noise subspaces, where we design a novel objective function in the form of the trace of density matrices product. Finally, a quantum labeling operation is proposed for the direction of arrival estimation of signal.

\begin{acknowledgments}
This work was supported by the National Science Foundation of China (No. 61871111 and No.
61960206005).
\end{acknowledgments}

\appendix
\section{NUMERICAL SIMULATIONS}
\label{appendix_1}
We define two isometry operators ${\rm M}$ and ${\rm N}$ so that 
\begin{equation}
\begin{array}{l}
{\rm M}{\rm{ = }}\sum\nolimits_{i = 0}^{Q - 1} {\left| i \right\rangle \left| {\mathop A\nolimits_i } \right\rangle } \left\langle i \right|:\left| i \right\rangle  \to \left| i \right\rangle \left| {\mathop A\nolimits_i } \right\rangle \\
 = \frac{1}{{\mathop {\left\| {\mathop A\nolimits_i } \right\|}\nolimits_2 }}\sum\nolimits_{j = 0}^{\mathop M\nolimits^2  - 1} {\mathop A\nolimits_{ij} \left| i \right\rangle \left| j \right\rangle } 
\end{array}
\end{equation}
\begin{equation}
\begin{array}{l}
{\rm N}{\rm{ = }}\sum\nolimits_{j = 0}^{\mathop M\nolimits^2  - 1} {\left| {\mathop {\left\| A \right\|}\nolimits_F } \right\rangle } \left| j \right\rangle \left\langle j \right|:\left| j \right\rangle  \to \left| {\mathop {\left\| A \right\|}\nolimits_F } \right\rangle \left| j \right\rangle \\
 = \frac{1}{{\mathop {\left\| A \right\|}\nolimits_F }}\sum\nolimits_{i = 0}^{Q - 1} {\mathop {\left\| {\mathop A\nolimits_i } \right\|}\nolimits_2 \left| i \right\rangle \left| j \right\rangle } 
\end{array}
\end{equation}
Similarly, we can obtain 
\begin{equation}
{\rm M}^H {\rm N} = \frac{1}{{\mathop {\left\| A \right\|}\nolimits_F }}\sum\nolimits_{i = 0}^{Q - 1} {\sum\nolimits_{j = 0}^{\mathop M\nolimits^2  - 1} {\mathop A\nolimits_{ij} \left| i \right\rangle \left\langle j \right|} }  = \frac{A}{{\mathop {\left\| A \right\|}\nolimits_F }}
\end{equation}
and ${\rm M}^H {\rm M} = \mathop I\nolimits_Q $ and ${\rm N}^H {\rm N} = I_{ M^2 } $.
Let the unitary transformation $W$ be
\begin{equation}
W = \left( {2{\rm N}{\rm N}^H  - I} \right)\left( {2{\rm M} {\rm M}^H  - I} \right)
\end{equation}
where $W$ can be efficiently implemented  with the complexity ${\rm{O}}\left( {{\rm{poly}}\left( {\log MQ} \right)} \right)$ based on unitaries $ U_{\rm M} $ and $U_{\rm N} $. Then, we perform $W$ on ${\rm N}\left| { v_i } \right\rangle $ and ${\rm M}\left| { u_i } \right\rangle $ as
\begin{equation}
\begin{array}{l}
\left( {2{\rm N} {\rm N}^H  - I} \right)\left( {2{\rm M}{\rm M}^H  - I} \right){\rm M}\left| {\mathop u\nolimits_i } \right\rangle \\
 = \left( {2{\rm N}{\rm N}^H  - I} \right){\rm M}\left| { u_i } \right\rangle  = 2{\rm N}{\rm N}^H {\rm M}\left| { u_i } \right\rangle  - {\rm M}\left| { u_i } \right\rangle \\
 = 2{\rm N}\frac{{A^H }}{{ {\left\| A \right\|}_F }}\left| { u_i } \right\rangle  - {\rm M}\left| { u_i } \right\rangle  = 2\frac{{\sigma_i {\rm N}\left| {v_i } \right\rangle }}{{ {\left\| A \right\|}_F }} - {\rm M}\left| {u_i } \right\rangle 
\end{array}
\end{equation}
\begin{equation}
\begin{array}{l}
\left( {2{\rm N}\mathop {\rm N}\nolimits^H  - I} \right)\left( {2{\rm M}\mathop {\rm M}\nolimits^H  - I} \right){\rm N}\left| {\mathop v\nolimits_i } \right\rangle \\
 = \left( {2{\rm N}\mathop {\rm N}\nolimits^H  - I} \right)\left( {2{\rm M}\mathop {\rm M}\nolimits^H {\rm N}\left| {\mathop v\nolimits_i } \right\rangle  - {\rm N}\left| {\mathop v\nolimits_i } \right\rangle } \right)\\
 = \left( {2{\rm N}\mathop {\rm N}\nolimits^H  - I} \right)\left( {2{\rm M}\frac{A}{{\mathop {\left\| A \right\|}\nolimits_F }}\left| {\mathop v\nolimits_i } \right\rangle  - {\rm N}\left| {\mathop v\nolimits_i } \right\rangle } \right) = \\
\left( {2{\rm N}\mathop {\rm N}\nolimits^H  - I} \right)\left( {2{\rm M}\frac{{\mathop \sigma \nolimits_i \left| {\mathop u\nolimits_i } \right\rangle }}{{\mathop {\left\| A \right\|}\nolimits_F }} - {\rm N}\left| {\mathop v\nolimits_i } \right\rangle } \right) = \\
4{\rm N}\frac{{\mathop A\nolimits^H }}{{\mathop {\left\| A \right\|}\nolimits_F }}\frac{{\mathop \sigma \nolimits_i \left| {\mathop u\nolimits_i } \right\rangle }}{{\mathop {\left\| A \right\|}\nolimits_F }} - 2{\rm N}\left| {\mathop v\nolimits_i } \right\rangle  - 2{\rm M}\frac{{\mathop \sigma \nolimits_i \left| {\mathop u\nolimits_i } \right\rangle }}{{\mathop {\left\| A \right\|}\nolimits_F }} + {\rm N}\left| {\mathop v\nolimits_i } \right\rangle  = \\
4{\rm N}\frac{{\mathop \sigma \nolimits_i^2 \left| {\mathop v\nolimits_i } \right\rangle }}{{\mathop {\left\| A \right\|}\nolimits_F^2 }} - 2{\rm M}\frac{{\mathop \sigma \nolimits_i \left| {\mathop u\nolimits_i } \right\rangle }}{{\mathop {\left\| A \right\|}\nolimits_F }} - {\rm N}\left| {\mathop v\nolimits_i } \right\rangle  = \\
\left( {\frac{{\mathop {4\sigma }\nolimits_i^2 }}{{\mathop {\left\| A \right\|}\nolimits_F^2 }} - 1} \right){\rm N}\left| {\mathop v\nolimits_i } \right\rangle  - 2{\rm M}\frac{{\mathop \sigma \nolimits_i \left| {\mathop u\nolimits_i } \right\rangle }}{{\mathop {\left\| A \right\|}\nolimits_F }}
\end{array}
\end{equation}
where $\left| {\mathop u\nolimits_i } \right\rangle $ and $\left| {\mathop v\nolimits_i } \right\rangle $ are the left and right singular vectors of $A$, and ${\mathop \sigma \nolimits_i }$ is the corresponding singular value. Obviously, the subspace $\left\{ {{\rm N}\left| {\mathop v\nolimits_i } \right\rangle ,{\rm M}\left| {\mathop u\nolimits_i } \right\rangle } \right\}$ is invariant under $W$, and ${{\rm M}\left| {\mathop u\nolimits_i } \right\rangle }$ and ${{\rm N}\left| {\mathop v\nolimits_i } \right\rangle }$ are lineary independent. Therefore, we apply the Gram-Schmidt process on ${{\rm M}\left| {\mathop u\nolimits_i } \right\rangle }$ and ${{\rm N}\left| {\mathop v\nolimits_i } \right\rangle }$ to obtain the orthogonal states $\left| {\mathop \varpi \nolimits_i^1 } \right\rangle $ and $\left| {\mathop \varpi \nolimits_i^2 } \right\rangle $ as follows
\begin{equation}
\begin{array}{l}
\left| {\mathop \varpi \nolimits_i^1 } \right\rangle  = {\rm M}\left| {\mathop u\nolimits_i } \right\rangle \\
\left| {\mathop \varpi \nolimits_i^2 } \right\rangle  = \frac{{{\rm N}\left| {\mathop v\nolimits_i } \right\rangle  - \left\langle {\mathop v\nolimits_i } \right|\mathop {\rm N}\nolimits^H {\rm M}\left| {\mathop u\nolimits_i } \right\rangle {\rm M}\left| {\mathop u\nolimits_i } \right\rangle }}{{\mathop {\left\| {{\rm N}\left| {\mathop v\nolimits_i } \right\rangle  - \left\langle {\mathop v\nolimits_i } \right|\mathop {\rm N}\nolimits^H {\rm M}\left| {\mathop u\nolimits_i } \right\rangle {\rm M}\left| {\mathop u\nolimits_i } \right\rangle } \right\|}\nolimits_2 }}\\
 = \frac{{{\rm N}\left| {\mathop v\nolimits_i } \right\rangle  - {\raise0.7ex\hbox{${\mathop \sigma \nolimits_i }$} \!\mathord{\left/
 {\vphantom {{\mathop \sigma \nolimits_i } {\mathop {\left\| A \right\|}\nolimits_F }}}\right.\kern-\nulldelimiterspace}
\!\lower0.7ex\hbox{${\mathop {\left\| A \right\|}\nolimits_F }$}}{\rm M}\left| {\mathop u\nolimits_i } \right\rangle }}{{\mathop {\left\| {{\rm N}\left| {\mathop v\nolimits_i } \right\rangle  - {\raise0.7ex\hbox{${\mathop \sigma \nolimits_i }$} \!\mathord{\left/
 {\vphantom {{\mathop \sigma \nolimits_i } {\mathop {\left\| A \right\|}\nolimits_F }}}\right.\kern-\nulldelimiterspace}
\!\lower0.7ex\hbox{${\mathop {\left\| A \right\|}\nolimits_F }$}}{\rm M}\left| {\mathop u\nolimits_i } \right\rangle } \right\|}\nolimits_2 }}
\end{array}
\end{equation}
where ${\mathop {\left\| {{\rm N}\left| {\mathop v\nolimits_i } \right\rangle  - {\raise0.7ex\hbox{${\mathop \sigma \nolimits_i }$} \!\mathord{\left/
 {\vphantom {{\mathop \sigma \nolimits_i } {\mathop {\left\| A \right\|}\nolimits_F }}}\right.\kern-\nulldelimiterspace}
\!\lower0.7ex\hbox{${\mathop {\left\| A \right\|}\nolimits_F }$}}{\rm M}\left| {\mathop u\nolimits_i } \right\rangle } \right\|}\nolimits_2 }$
can be deduced as
\begin{equation}
\begin{array}{l}
\mathop {\left\| {{\rm N}\left| {\mathop v\nolimits_i } \right\rangle  - {\raise0.7ex\hbox{${\mathop \sigma \nolimits_i }$} \!\mathord{\left/
 {\vphantom {{\mathop \sigma \nolimits_i } {\mathop {\left\| A \right\|}\nolimits_F }}}\right.\kern-\nulldelimiterspace}
\!\lower0.7ex\hbox{${\mathop {\left\| A \right\|}\nolimits_F }$}}{\rm M}\left| {\mathop u\nolimits_i } \right\rangle } \right\|}\nolimits_2 \\
{\rm{ = }}\sqrt {\mathop {\left( {{\rm N}\left| {\mathop v\nolimits_i } \right\rangle  - {\raise0.7ex\hbox{${\mathop \sigma \nolimits_i }$} \!\mathord{\left/
 {\vphantom {{\mathop \sigma \nolimits_i } {\mathop {\left\| A \right\|}\nolimits_F }}}\right.\kern-\nulldelimiterspace}
\!\lower0.7ex\hbox{${\mathop {\left\| A \right\|}\nolimits_F }$}}{\rm M}\left| {\mathop u\nolimits_i } \right\rangle } \right)}\nolimits^H \left( {{\rm N}\left| {\mathop v\nolimits_i } \right\rangle  - {\raise0.7ex\hbox{${\mathop \sigma \nolimits_i }$} \!\mathord{\left/
 {\vphantom {{\mathop \sigma \nolimits_i } {\mathop {\left\| A \right\|}\nolimits_F }}}\right.\kern-\nulldelimiterspace}
\!\lower0.7ex\hbox{${\mathop {\left\| A \right\|}\nolimits_F }$}}{\rm M}\left| {\mathop u\nolimits_i } \right\rangle } \right)} \\
 = \sqrt {1 - \mathop {\left( {{\raise0.7ex\hbox{${\mathop \sigma \nolimits_i }$} \!\mathord{\left/
 {\vphantom {{\mathop \sigma \nolimits_i } {\mathop {\left\| A \right\|}\nolimits_F }}}\right.\kern-\nulldelimiterspace}
\!\lower0.7ex\hbox{${\mathop {\left\| A \right\|}\nolimits_F }$}}} \right)}\nolimits^2  - \mathop {\left( {{\raise0.7ex\hbox{${\mathop \sigma \nolimits_i }$} \!\mathord{\left/
 {\vphantom {{\mathop \sigma \nolimits_i } {\mathop {\left\| A \right\|}\nolimits_F }}}\right.\kern-\nulldelimiterspace}
\!\lower0.7ex\hbox{${\mathop {\left\| A \right\|}\nolimits_F }$}}} \right)}\nolimits^2  + \mathop {\left( {{\raise0.7ex\hbox{${\mathop \sigma \nolimits_i }$} \!\mathord{\left/
 {\vphantom {{\mathop \sigma \nolimits_i } {\mathop {\left\| A \right\|}\nolimits_F }}}\right.\kern-\nulldelimiterspace}
\!\lower0.7ex\hbox{${\mathop {\left\| A \right\|}\nolimits_F }$}}} \right)}\nolimits^2 } \\
 = \sqrt {1 - \mathop {\left( {{\raise0.7ex\hbox{${\mathop \sigma \nolimits_i }$} \!\mathord{\left/
 {\vphantom {{\mathop \sigma \nolimits_i } {\mathop {\left\| A \right\|}\nolimits_F }}}\right.\kern-\nulldelimiterspace}
\!\lower0.7ex\hbox{${\mathop {\left\| A \right\|}\nolimits_F }$}}} \right)}\nolimits^2 } 
\end{array}
\end{equation}
Thus, we have
\begin{equation}
\begin{array}{l}
W\left| {\mathop \varpi \nolimits_i^1 } \right\rangle  = W{\rm M}\left| {\mathop u\nolimits_i } \right\rangle  = 2\frac{{\mathop \sigma \nolimits_i }}{{\mathop {\left\| A \right\|}\nolimits_F }}{\rm N}\left| {\mathop v\nolimits_i } \right\rangle  - \left| {\mathop \varpi \nolimits_i^1 } \right\rangle  = \\
2\frac{{\mathop \sigma \nolimits_i }}{{\mathop {\left\| A \right\|}\nolimits_F }}\left( {\sqrt {1 - \mathop {\left( {\frac{{\mathop \sigma \nolimits_i }}{{\mathop {\left\| A \right\|}\nolimits_F }}} \right)}\nolimits^2 } \left| {\mathop \varpi \nolimits_i^2 } \right\rangle  + \frac{{\mathop \sigma \nolimits_i }}{{\mathop {\left\| A \right\|}\nolimits_F }}\left| {\mathop \varpi \nolimits_i^1 } \right\rangle } \right) - \left| {\mathop \varpi \nolimits_i^1 } \right\rangle  = \\
2\frac{{\mathop \sigma \nolimits_i }}{{\mathop {\left\| A \right\|}\nolimits_F }}\sqrt {1 - \mathop {\left( {\frac{{\mathop \sigma \nolimits_i }}{{\mathop {\left\| A \right\|}\nolimits_F }}} \right)}\nolimits^2 } \left| {\mathop \varpi \nolimits_i^2 } \right\rangle  + \left( {2\mathop {\left( {\frac{{\mathop \sigma \nolimits_i }}{{\mathop {\left\| A \right\|}\nolimits_F }}} \right)}\nolimits^2  - 1} \right)\left| {\mathop \varpi \nolimits_i^1 } \right\rangle 
\end{array}
\end{equation}
\begin{equation}
\begin{array}{l}
W\left| {\mathop \varpi \nolimits_i^2 } \right\rangle  = \frac{{W{\rm N}\left| {\mathop v\nolimits_i } \right\rangle  - \frac{{\mathop \sigma \nolimits_i }}{{\mathop {\left\| A \right\|}\nolimits_F }}W{\rm M}\left| {\mathop u\nolimits_i } \right\rangle }}{{\sqrt {1 - \mathop {\left( {\frac{{\mathop \sigma \nolimits_i }}{{\mathop {\left\| A \right\|}\nolimits_F }}} \right)}\nolimits^2 } }} = \\
 - \frac{{\mathop {2\sigma }\nolimits_i }}{{\mathop {\left\| A \right\|}\nolimits_F }}\sqrt {1 - \mathop {\left( {\frac{{\mathop \sigma \nolimits_i }}{{\mathop {\left\| A \right\|}\nolimits_F }}} \right)}\nolimits^2 } \left| {\mathop \varpi \nolimits_i^1 } \right\rangle  + \left( {2\mathop {\left( {\frac{{\mathop \sigma \nolimits_i }}{{\mathop {\left\| A \right\|}\nolimits_F }}} \right)}\nolimits^2  - 1} \right)\left| {\mathop \varpi \nolimits_i^2 } \right\rangle 
\end{array}
\end{equation}
Then, we define ${2\mathop {\left( {{\raise0.7ex\hbox{${\mathop \sigma \nolimits_i }$} \!\mathord{\left/
 {\vphantom {{\mathop \sigma \nolimits_i } {\mathop {\left\| A \right\|}\nolimits_F }}}\right.\kern-\nulldelimiterspace}
\!\lower0.7ex\hbox{${\mathop {\left\| A \right\|}\nolimits_F }$}}} \right)}\nolimits^2  - 1}$ to be $\cos \mathop \chi \nolimits_i $, that is $\cos {\raise0.7ex\hbox{${\mathop \chi \nolimits_i }$} \!\mathord{\left/
 {\vphantom {{\mathop \chi \nolimits_i } 2}}\right.\kern-\nulldelimiterspace}
\!\lower0.7ex\hbox{$2$}} = {\raise0.7ex\hbox{${\mathop \sigma \nolimits_i }$} \!\mathord{\left/
 {\vphantom {{\mathop \sigma \nolimits_i } {\mathop {\left\| A \right\|}\nolimits_F }}}\right.\kern-\nulldelimiterspace}
\!\lower0.7ex\hbox{${\mathop {\left\| A \right\|}\nolimits_F }$}}$, we can represent the Eqs. (A9) and (A10) as
\begin{equation}
W\left| {\mathop \varpi \nolimits_i^1 } \right\rangle  = \sin \mathop \chi \nolimits_i \left| {\mathop \varpi \nolimits_i^2 } \right\rangle  + \cos \mathop \chi \nolimits_i \left| {\mathop \varpi \nolimits_i^1 } \right\rangle 
\end{equation}
\begin{equation}
W\left| {\mathop \varpi \nolimits_i^2 } \right\rangle  =  - \sin \mathop \chi \nolimits_i \left| {\mathop \varpi \nolimits_i^1 } \right\rangle  + \cos \mathop \chi \nolimits_i \left| {\mathop \varpi \nolimits_i^2 } \right\rangle 
\end{equation}
Moreover, the matrix form of $W$ in basis $\left| {\mathop \varpi \nolimits_i^1 } \right\rangle $ and $\left| {\mathop \varpi \nolimits_i^2 } \right\rangle $ can be denoted as
\begin{equation}
\left( {\begin{array}{*{20}{c}}
{\cos \mathop \chi \nolimits_i }&{\sin \mathop \chi \nolimits_i }\\
{ - \sin \mathop \chi \nolimits_i }&{\cos \mathop \chi \nolimits_i }
\end{array}} \right) = \mathop e\nolimits^{i\mathop \chi \nolimits_i \mathop \sigma \nolimits_y } 
\end{equation}
where ${\mathop \sigma \nolimits_y }$ is a Pauli matrix in basis $\left| {\mathop \varpi \nolimits_i^1 } \right\rangle $ and $\left| {\mathop \varpi \nolimits_i^2 } \right\rangle $ . Therefore, the eigenvalues of $W$ are $\mathop e\nolimits^{i\mathop { \pm \chi }\nolimits_i } $ and the corresponding eigenvectors are $\left| {\mathop \lambda \nolimits_i^ \pm  } \right\rangle  = \frac{1}{{\sqrt 2 }}\left( {\left| {\mathop \varpi \nolimits_i^1 } \right\rangle  \pm i\left| {\mathop \varpi \nolimits_i^2 } \right\rangle } \right)$. Then, ${{\rm M}\left| {\mathop u\nolimits_i } \right\rangle }$ and ${{\rm N}\left| {\mathop v\nolimits_i } \right\rangle }$ can be reformulated as
\begin{equation}
{\rm M}\left| {\mathop u\nolimits_i } \right\rangle  = \left| {\mathop \varpi \nolimits_i^1 } \right\rangle  = \frac{1}{{\sqrt 2 }}\left( {\left| {\mathop \lambda \nolimits_i^ +  } \right\rangle  + \left| {\mathop \lambda \nolimits_i^ -  } \right\rangle } \right)
\end{equation}
\begin{equation}
\begin{array}{l}
{\rm N}\left| {\mathop v\nolimits_i } \right\rangle  = \sin \frac{{\mathop \chi \nolimits_i }}{2}\left| {\mathop \varpi \nolimits_i^2 } \right\rangle  + \cos \frac{{\mathop \chi \nolimits_i }}{2}\left| {\mathop \varpi \nolimits_i^1 } \right\rangle  = \\
\sin \frac{{\mathop \chi \nolimits_i }}{2}\frac{{ - i}}{{\sqrt 2 }}\left( {\left| {\mathop \lambda \nolimits_i^ +  } \right\rangle  - \left| {\mathop \lambda \nolimits_i^ -  } \right\rangle } \right) + \cos \frac{{\mathop \chi \nolimits_i }}{2}\frac{1}{{\sqrt 2 }}\left( {\left| {\mathop \lambda \nolimits_i^ +  } \right\rangle  + \left| {\mathop \lambda \nolimits_i^ -  } \right\rangle } \right)\\
 = \frac{1}{{\sqrt 2 }}\left( {\mathop e\nolimits^{ - i\frac{{\mathop \chi \nolimits_i }}{2}} \left| {\mathop \lambda \nolimits_i^ +  } \right\rangle  + \mathop e\nolimits^{i\frac{{\mathop \chi \nolimits_i }}{2}} \left| {\mathop \lambda \nolimits_i^ -  } \right\rangle } \right)
\end{array}
\end{equation}
Therefore, the detailed implementation of the transformation ${\rm T}$ can be shown as follows:
\\
(1) Given a vector $\left| y \right\rangle  = \sum {\mathop \beta \nolimits_i } \left| {\mathop u\nolimits_i } \right\rangle $, we prepare the state 
\begin{equation}
\begin{array}{l}
\left| \vartheta  \right\rangle  = {\rm M}\left| y \right\rangle  = \sum {\mathop \beta \nolimits_i } {\rm M}\left| {\mathop u\nolimits_i } \right\rangle  = \\
\sum {\frac{{\mathop \beta \nolimits_i }}{{\sqrt 2 }}\left( {\left| {\mathop \lambda \nolimits_i^ +  } \right\rangle  + \left| {\mathop \lambda \nolimits_i^ -  } \right\rangle } \right)} 
\end{array}
\end{equation}
(2) Perform the phase estimation algorithm $\mathop U\nolimits_{PE} \left( W \right)$,  we have the state
\begin{equation}
\mathop U\nolimits_{PE} \left( W \right):\left| \vartheta  \right\rangle  \to \sum {\frac{{\mathop \beta \nolimits_i }}{{\sqrt 2 }}\left( {\left| {\mathop \lambda \nolimits_i^ +  } \right\rangle \left| {\mathop \chi \nolimits_i } \right\rangle  + \left| {\mathop \lambda \nolimits_i^ -  } \right\rangle \left| {\mathop { - \chi }\nolimits_i } \right\rangle } \right)} 
\end{equation}
(3) Apply the controlled phase rotation operation, the state is transformed into
\begin{equation}
\left| \vartheta  \right\rangle  = \sum {\frac{{\mathop \beta \nolimits_i }}{{\sqrt 2 }}\left( {\mathop e\nolimits^{ - i\frac{{\mathop \chi \nolimits_i }}{2}} \left| {\mathop \lambda \nolimits_i^ +  } \right\rangle \left| {\mathop \chi \nolimits_i } \right\rangle  + \mathop e\nolimits^{i\frac{{\mathop \chi \nolimits_i }}{2}} \left| {\mathop \lambda \nolimits_i^ -  } \right\rangle \left| {\mathop { - \chi }\nolimits_i } \right\rangle } \right)} 
\end{equation}
(4) Undo the phase estimation algorithm, we have the state
\begin{equation}
\left| \vartheta  \right\rangle  = \sum {\frac{{\mathop \beta \nolimits_i }}{{\sqrt 2 }}\left( {\mathop e\nolimits^{ - i\frac{{\mathop \chi \nolimits_i }}{2}} \left| {\mathop \lambda \nolimits_i^ +  } \right\rangle  + \mathop e\nolimits^{i\frac{{\mathop \chi \nolimits_i }}{2}} \left| {\mathop \lambda \nolimits_i^ -  } \right\rangle } \right)}  = \sum {\mathop \beta \nolimits_i } {\rm N}\left| {\mathop v\nolimits_i } \right\rangle 
\end{equation}
(5) Perform the inverse operation $\mathop U\nolimits_{\rm N}^{ - 1} $, we can obtain
\begin{equation}
\mathop U\nolimits_{\rm N}^{ - 1} {\rm{ = }}\mathop U\nolimits_{\rm N}^H :\left| \vartheta  \right\rangle  \to \sum {\mathop \beta \nolimits_i } \left| {\mathop v\nolimits_i } \right\rangle 
\end{equation}

%\newpage
%\nocite{*}
\bibliography{bib3}

\begin{thebibliography}{55}%
\makeatletter
\providecommand \@ifxundefined [1]{%
 \@ifx{#1\undefined}
}%
\providecommand \@ifnum [1]{%
 \ifnum #1\expandafter \@firstoftwo
 \else \expandafter \@secondoftwo
 \fi
}%
\providecommand \@ifx [1]{%
 \ifx #1\expandafter \@firstoftwo
 \else \expandafter \@secondoftwo
 \fi
}%
\providecommand \natexlab [1]{#1}%
\providecommand \enquote  [1]{``#1''}%
\providecommand \bibnamefont  [1]{#1}%
\providecommand \bibfnamefont [1]{#1}%
\providecommand \citenamefont [1]{#1}%
\providecommand \href@noop [0]{\@secondoftwo}%
\providecommand \href [0]{\begingroup \@sanitize@url \@href}%
\providecommand \@href[1]{\@@startlink{#1}\@@href}%
\providecommand \@@href[1]{\endgroup#1\@@endlink}%
\providecommand \@sanitize@url [0]{\catcode `\\12\catcode `\$12\catcode
  `\&12\catcode `\#12\catcode `\^12\catcode `\_12\catcode `\%12\relax}%
\providecommand \@@startlink[1]{}%
\providecommand \@@endlink[0]{}%
\providecommand \url  [0]{\begingroup\@sanitize@url \@url }%
\providecommand \@url [1]{\endgroup\@href {#1}{\urlprefix }}%
\providecommand \urlprefix  [0]{URL }%
\providecommand \Eprint [0]{\href }%
\providecommand \doibase [0]{http://dx.doi.org/}%
\providecommand \selectlanguage [0]{\@gobble}%
\providecommand \bibinfo  [0]{\@secondoftwo}%
\providecommand \bibfield  [0]{\@secondoftwo}%
\providecommand \translation [1]{[#1]}%
\providecommand \BibitemOpen [0]{}%
\providecommand \bibitemStop [0]{}%
\providecommand \bibitemNoStop [0]{.\EOS\space}%
\providecommand \EOS [0]{\spacefactor3000\relax}%
\providecommand \BibitemShut  [1]{\csname bibitem#1\endcsname}%
\let\auto@bib@innerbib\@empty
%</preamble>
\bibitem [{\citenamefont {Boyer}\ and\ \citenamefont {Bouleux}(2008)}]{1}%
  \BibitemOpen
  \bibfield  {author} {\bibinfo {author} {\bibfnamefont {R.}~\bibnamefont
  {Boyer}}\ and\ \bibinfo {author} {\bibfnamefont {G.}~\bibnamefont
  {Bouleux}},\ }\href@noop {} {\bibfield  {journal} {\bibinfo  {journal} {IEEE
  Transactions on Signal Processing}\ }\textbf {\bibinfo {volume} {56}},\
  \bibinfo {pages} {1374} (\bibinfo {year} {2008})}\BibitemShut {NoStop}%
\bibitem [{\citenamefont {Zheng}\ and\ \citenamefont {Kaveh}(2013)}]{2}%
  \BibitemOpen
  \bibfield  {author} {\bibinfo {author} {\bibfnamefont {J.}~\bibnamefont
  {Zheng}}\ and\ \bibinfo {author} {\bibfnamefont {M.}~\bibnamefont {Kaveh}},\
  }\href@noop {} {\bibfield  {journal} {\bibinfo  {journal} {IEEE Transactions
  on Signal Processing}\ }\textbf {\bibinfo {volume} {61}},\ \bibinfo {pages}
  {2767} (\bibinfo {year} {2013})}\BibitemShut {NoStop}%
\bibitem [{\citenamefont {Duofang}\ \emph {et~al.}(2008)\citenamefont
  {Duofang}, \citenamefont {Baixiao},\ and\ \citenamefont {Guodong}}]{3}%
  \BibitemOpen
  \bibfield  {author} {\bibinfo {author} {\bibfnamefont {C.}~\bibnamefont
  {Duofang}}, \bibinfo {author} {\bibfnamefont {C.}~\bibnamefont {Baixiao}}, \
  and\ \bibinfo {author} {\bibfnamefont {Q.}~\bibnamefont {Guodong}},\
  }\href@noop {} {\bibfield  {journal} {\bibinfo  {journal} {Electronics
  Letters}\ }\textbf {\bibinfo {volume} {44}},\ \bibinfo {pages} {770}
  (\bibinfo {year} {2008})}\BibitemShut {NoStop}%
\bibitem [{\citenamefont {Jinli}\ \emph {et~al.}(2008)\citenamefont {Jinli},
  \citenamefont {Hong},\ and\ \citenamefont {Weimin}}]{4}%
  \BibitemOpen
  \bibfield  {author} {\bibinfo {author} {\bibfnamefont {C.}~\bibnamefont
  {Jinli}}, \bibinfo {author} {\bibfnamefont {G.}~\bibnamefont {Hong}}, \ and\
  \bibinfo {author} {\bibfnamefont {S.}~\bibnamefont {Weimin}},\ }\href@noop {}
  {\bibfield  {journal} {\bibinfo  {journal} {Electronics Letters}\ }\textbf
  {\bibinfo {volume} {44}},\ \bibinfo {pages} {1422} (\bibinfo {year}
  {2008})}\BibitemShut {NoStop}%
\bibitem [{\citenamefont {Zhang}\ and\ \citenamefont {Xu}(2011)}]{5}%
  \BibitemOpen
  \bibfield  {author} {\bibinfo {author} {\bibfnamefont {X.}~\bibnamefont
  {Zhang}}\ and\ \bibinfo {author} {\bibfnamefont {D.}~\bibnamefont {Xu}},\
  }\href@noop {} {\bibfield  {journal} {\bibinfo  {journal} {Electronics
  Letters}\ }\textbf {\bibinfo {volume} {47}},\ \bibinfo {pages} {283}
  (\bibinfo {year} {2011})}\BibitemShut {NoStop}%
\bibitem [{\citenamefont {Zheng}\ \emph {et~al.}(2012)\citenamefont {Zheng},
  \citenamefont {Chen},\ and\ \citenamefont {Yang}}]{6}%
  \BibitemOpen
  \bibfield  {author} {\bibinfo {author} {\bibfnamefont {G.}~\bibnamefont
  {Zheng}}, \bibinfo {author} {\bibfnamefont {B.}~\bibnamefont {Chen}}, \ and\
  \bibinfo {author} {\bibfnamefont {M.}~\bibnamefont {Yang}},\ }\href@noop {}
  {\bibfield  {journal} {\bibinfo  {journal} {Electronics Letters}\ }\textbf
  {\bibinfo {volume} {48}},\ \bibinfo {pages} {179} (\bibinfo {year}
  {2012})}\BibitemShut {NoStop}%
\bibitem [{\citenamefont {Liao}(2018)}]{7}%
  \BibitemOpen
  \bibfield  {author} {\bibinfo {author} {\bibfnamefont {B.}~\bibnamefont
  {Liao}},\ }\href@noop {} {\bibfield  {journal} {\bibinfo  {journal} {IEEE
  Transactions on Aerospace and Electronic Systems}\ }\textbf {\bibinfo
  {volume} {54}},\ \bibinfo {pages} {2091} (\bibinfo {year}
  {2018})}\BibitemShut {NoStop}%
\bibitem [{\citenamefont {SCHMIDT}(1981)}]{8}%
  \BibitemOpen
  \bibfield  {author} {\bibinfo {author} {\bibfnamefont {R.}~\bibnamefont
  {SCHMIDT}},\ }\href@noop {} {\bibfield  {journal} {\bibinfo  {journal} {Ph.
  D. Dissertation. Stanford Univ.}\ } (\bibinfo {year} {1981})}\BibitemShut
  {NoStop}%
\bibitem [{\citenamefont {Larsson}\ \emph {et~al.}(2014)\citenamefont
  {Larsson}, \citenamefont {Edfors}, \citenamefont {Tufvesson},\ and\
  \citenamefont {Marzetta}}]{9}%
  \BibitemOpen
  \bibfield  {author} {\bibinfo {author} {\bibfnamefont {E.~G.}\ \bibnamefont
  {Larsson}}, \bibinfo {author} {\bibfnamefont {O.}~\bibnamefont {Edfors}},
  \bibinfo {author} {\bibfnamefont {F.}~\bibnamefont {Tufvesson}}, \ and\
  \bibinfo {author} {\bibfnamefont {T.~L.}\ \bibnamefont {Marzetta}},\
  }\href@noop {} {\bibfield  {journal} {\bibinfo  {journal} {IEEE
  communications magazine}\ }\textbf {\bibinfo {volume} {52}},\ \bibinfo
  {pages} {186} (\bibinfo {year} {2014})}\BibitemShut {NoStop}%
\bibitem [{\citenamefont {Liang}\ \emph {et~al.}(2014)\citenamefont {Liang},
  \citenamefont {Xu},\ and\ \citenamefont {Dong}}]{10}%
  \BibitemOpen
  \bibfield  {author} {\bibinfo {author} {\bibfnamefont {L.}~\bibnamefont
  {Liang}}, \bibinfo {author} {\bibfnamefont {W.}~\bibnamefont {Xu}}, \ and\
  \bibinfo {author} {\bibfnamefont {X.}~\bibnamefont {Dong}},\ }\href@noop {}
  {\bibfield  {journal} {\bibinfo  {journal} {IEEE Wireless Communications
  Letters}\ }\textbf {\bibinfo {volume} {3}},\ \bibinfo {pages} {653} (\bibinfo
  {year} {2014})}\BibitemShut {NoStop}%
\bibitem [{\citenamefont {El~Ayach}\ \emph {et~al.}(2014)\citenamefont
  {El~Ayach}, \citenamefont {Rajagopal}, \citenamefont {Abu-Surra},
  \citenamefont {Pi},\ and\ \citenamefont {Heath}}]{11}%
  \BibitemOpen
  \bibfield  {author} {\bibinfo {author} {\bibfnamefont {O.}~\bibnamefont
  {El~Ayach}}, \bibinfo {author} {\bibfnamefont {S.}~\bibnamefont {Rajagopal}},
  \bibinfo {author} {\bibfnamefont {S.}~\bibnamefont {Abu-Surra}}, \bibinfo
  {author} {\bibfnamefont {Z.}~\bibnamefont {Pi}}, \ and\ \bibinfo {author}
  {\bibfnamefont {R.~W.}\ \bibnamefont {Heath}},\ }\href@noop {} {\bibfield
  {journal} {\bibinfo  {journal} {IEEE transactions on wireless
  communications}\ }\textbf {\bibinfo {volume} {13}},\ \bibinfo {pages} {1499}
  (\bibinfo {year} {2014})}\BibitemShut {NoStop}%
\bibitem [{\citenamefont {Venkateswaran}\ and\ \citenamefont {van~der
  Veen}(2010)}]{12}%
  \BibitemOpen
  \bibfield  {author} {\bibinfo {author} {\bibfnamefont {V.}~\bibnamefont
  {Venkateswaran}}\ and\ \bibinfo {author} {\bibfnamefont {A.-J.}\ \bibnamefont
  {van~der Veen}},\ }\href@noop {} {\bibfield  {journal} {\bibinfo  {journal}
  {IEEE Transactions on Signal Processing}\ }\textbf {\bibinfo {volume} {58}},\
  \bibinfo {pages} {4131} (\bibinfo {year} {2010})}\BibitemShut {NoStop}%
\bibitem [{\citenamefont {Lin}\ and\ \citenamefont {Li}(2015)}]{13}%
  \BibitemOpen
  \bibfield  {author} {\bibinfo {author} {\bibfnamefont {C.}~\bibnamefont
  {Lin}}\ and\ \bibinfo {author} {\bibfnamefont {G.~Y.}\ \bibnamefont {Li}},\
  }\href@noop {} {\bibfield  {journal} {\bibinfo  {journal} {IEEE Transactions
  on Communications}\ }\textbf {\bibinfo {volume} {63}},\ \bibinfo {pages}
  {2985} (\bibinfo {year} {2015})}\BibitemShut {NoStop}%
\bibitem [{\citenamefont {Li}\ \emph {et~al.}(2020)\citenamefont {Li},
  \citenamefont {Liu}, \citenamefont {You}, \citenamefont {Wang}, \citenamefont
  {Duan},\ and\ \citenamefont {Li}}]{14}%
  \BibitemOpen
  \bibfield  {author} {\bibinfo {author} {\bibfnamefont {S.}~\bibnamefont
  {Li}}, \bibinfo {author} {\bibfnamefont {Y.}~\bibnamefont {Liu}}, \bibinfo
  {author} {\bibfnamefont {L.}~\bibnamefont {You}}, \bibinfo {author}
  {\bibfnamefont {W.}~\bibnamefont {Wang}}, \bibinfo {author} {\bibfnamefont
  {H.}~\bibnamefont {Duan}}, \ and\ \bibinfo {author} {\bibfnamefont
  {X.}~\bibnamefont {Li}},\ }\href@noop {} {\bibfield  {journal} {\bibinfo
  {journal} {IEEE Wireless Communications Letters}\ } (\bibinfo {year}
  {2020})}\BibitemShut {NoStop}%
\bibitem [{\citenamefont {Ekert}\ and\ \citenamefont {Jozsa}(1996)}]{15}%
  \BibitemOpen
  \bibfield  {author} {\bibinfo {author} {\bibfnamefont {A.}~\bibnamefont
  {Ekert}}\ and\ \bibinfo {author} {\bibfnamefont {R.}~\bibnamefont {Jozsa}},\
  }\href@noop {} {\bibfield  {journal} {\bibinfo  {journal} {Reviews of Modern
  Physics}\ }\textbf {\bibinfo {volume} {68}},\ \bibinfo {pages} {733}
  (\bibinfo {year} {1996})}\BibitemShut {NoStop}%
\bibitem [{\citenamefont {Grover}(1997)}]{16}%
  \BibitemOpen
  \bibfield  {author} {\bibinfo {author} {\bibfnamefont {L.~K.}\ \bibnamefont
  {Grover}},\ }\href@noop {} {\bibfield  {journal} {\bibinfo  {journal}
  {Physical review letters}\ }\textbf {\bibinfo {volume} {79}},\ \bibinfo
  {pages} {325} (\bibinfo {year} {1997})}\BibitemShut {NoStop}%
\bibitem [{\citenamefont {Harrow}\ \emph {et~al.}(2009)\citenamefont {Harrow},
  \citenamefont {Hassidim},\ and\ \citenamefont {Lloyd}}]{17}%
  \BibitemOpen
  \bibfield  {author} {\bibinfo {author} {\bibfnamefont {A.~W.}\ \bibnamefont
  {Harrow}}, \bibinfo {author} {\bibfnamefont {A.}~\bibnamefont {Hassidim}}, \
  and\ \bibinfo {author} {\bibfnamefont {S.}~\bibnamefont {Lloyd}},\
  }\href@noop {} {\bibfield  {journal} {\bibinfo  {journal} {Physical review
  letters}\ }\textbf {\bibinfo {volume} {103}},\ \bibinfo {pages} {150502}
  (\bibinfo {year} {2009})}\BibitemShut {NoStop}%
\bibitem [{\citenamefont {Wiebe}\ \emph {et~al.}(2012)\citenamefont {Wiebe},
  \citenamefont {Braun},\ and\ \citenamefont {Lloyd}}]{18}%
  \BibitemOpen
  \bibfield  {author} {\bibinfo {author} {\bibfnamefont {N.}~\bibnamefont
  {Wiebe}}, \bibinfo {author} {\bibfnamefont {D.}~\bibnamefont {Braun}}, \ and\
  \bibinfo {author} {\bibfnamefont {S.}~\bibnamefont {Lloyd}},\ }\href@noop {}
  {\bibfield  {journal} {\bibinfo  {journal} {Physical review letters}\
  }\textbf {\bibinfo {volume} {109}},\ \bibinfo {pages} {050505} (\bibinfo
  {year} {2012})}\BibitemShut {NoStop}%
\bibitem [{\citenamefont {Clader}\ \emph {et~al.}(2013)\citenamefont {Clader},
  \citenamefont {Jacobs},\ and\ \citenamefont {Sprouse}}]{19}%
  \BibitemOpen
  \bibfield  {author} {\bibinfo {author} {\bibfnamefont {B.~D.}\ \bibnamefont
  {Clader}}, \bibinfo {author} {\bibfnamefont {B.~C.}\ \bibnamefont {Jacobs}},
  \ and\ \bibinfo {author} {\bibfnamefont {C.~R.}\ \bibnamefont {Sprouse}},\
  }\href@noop {} {\bibfield  {journal} {\bibinfo  {journal} {Physical review
  letters}\ }\textbf {\bibinfo {volume} {110}},\ \bibinfo {pages} {250504}
  (\bibinfo {year} {2013})}\BibitemShut {NoStop}%
\bibitem [{\citenamefont {Lloyd}\ \emph {et~al.}(2013)\citenamefont {Lloyd},
  \citenamefont {Mohseni},\ and\ \citenamefont {Rebentrost}}]{20}%
  \BibitemOpen
  \bibfield  {author} {\bibinfo {author} {\bibfnamefont {S.}~\bibnamefont
  {Lloyd}}, \bibinfo {author} {\bibfnamefont {M.}~\bibnamefont {Mohseni}}, \
  and\ \bibinfo {author} {\bibfnamefont {P.}~\bibnamefont {Rebentrost}},\
  }\href@noop {} {\bibfield  {journal} {\bibinfo  {journal} {arXiv preprint
  arXiv:1307.0411}\ } (\bibinfo {year} {2013})}\BibitemShut {NoStop}%
\bibitem [{\citenamefont {Lloyd}\ \emph {et~al.}(2014)\citenamefont {Lloyd},
  \citenamefont {Mohseni},\ and\ \citenamefont {Rebentrost}}]{21}%
  \BibitemOpen
  \bibfield  {author} {\bibinfo {author} {\bibfnamefont {S.}~\bibnamefont
  {Lloyd}}, \bibinfo {author} {\bibfnamefont {M.}~\bibnamefont {Mohseni}}, \
  and\ \bibinfo {author} {\bibfnamefont {P.}~\bibnamefont {Rebentrost}},\
  }\href@noop {} {\bibfield  {journal} {\bibinfo  {journal} {Nature Physics}\
  }\textbf {\bibinfo {volume} {10}},\ \bibinfo {pages} {631} (\bibinfo {year}
  {2014})}\BibitemShut {NoStop}%
\bibitem [{\citenamefont {Rebentrost}\ \emph {et~al.}(2014)\citenamefont
  {Rebentrost}, \citenamefont {Mohseni},\ and\ \citenamefont {Lloyd}}]{22}%
  \BibitemOpen
  \bibfield  {author} {\bibinfo {author} {\bibfnamefont {P.}~\bibnamefont
  {Rebentrost}}, \bibinfo {author} {\bibfnamefont {M.}~\bibnamefont {Mohseni}},
  \ and\ \bibinfo {author} {\bibfnamefont {S.}~\bibnamefont {Lloyd}},\
  }\href@noop {} {\bibfield  {journal} {\bibinfo  {journal} {Physical review
  letters}\ }\textbf {\bibinfo {volume} {113}},\ \bibinfo {pages} {130503}
  (\bibinfo {year} {2014})}\BibitemShut {NoStop}%
\bibitem [{\citenamefont {Cong}\ and\ \citenamefont {Duan}(2016)}]{23}%
  \BibitemOpen
  \bibfield  {author} {\bibinfo {author} {\bibfnamefont {I.}~\bibnamefont
  {Cong}}\ and\ \bibinfo {author} {\bibfnamefont {L.}~\bibnamefont {Duan}},\
  }\href@noop {} {\bibfield  {journal} {\bibinfo  {journal} {New Journal of
  Physics}\ }\textbf {\bibinfo {volume} {18}},\ \bibinfo {pages} {073011}
  (\bibinfo {year} {2016})}\BibitemShut {NoStop}%
\bibitem [{\citenamefont {Kerenidis}\ and\ \citenamefont {Prakash}(2017)}]{24}%
  \BibitemOpen
  \bibfield  {author} {\bibinfo {author} {\bibfnamefont {I.}~\bibnamefont
  {Kerenidis}}\ and\ \bibinfo {author} {\bibfnamefont {A.}~\bibnamefont
  {Prakash}},\ }in\ \href {\doibase 10.4230/LIPIcs.ITCS.2017.49} {\emph
  {\bibinfo {booktitle} {8th Innovations in Theoretical Computer Science
  Conference (ITCS 2017)}}},\ \bibinfo {series} {Leibniz International
  Proceedings in Informatics (LIPIcs)}, Vol.~\bibinfo {volume} {67},\ \bibinfo
  {editor} {edited by\ \bibinfo {editor} {\bibfnamefont {C.~H.}\ \bibnamefont
  {Papadimitriou}}}\ (\bibinfo  {publisher} {Schloss Dagstuhl--Leibniz-Zentrum
  fuer Informatik},\ \bibinfo {address} {Dagstuhl, Germany},\ \bibinfo {year}
  {2017})\ pp.\ \bibinfo {pages} {49:1--49:21}\BibitemShut {NoStop}%
\bibitem [{\citenamefont {Kerenidis}\ and\ \citenamefont {Prakash}(2020)}]{25}%
  \BibitemOpen
  \bibfield  {author} {\bibinfo {author} {\bibfnamefont {I.}~\bibnamefont
  {Kerenidis}}\ and\ \bibinfo {author} {\bibfnamefont {A.}~\bibnamefont
  {Prakash}},\ }\href@noop {} {\bibfield  {journal} {\bibinfo  {journal}
  {Physical Review A}\ }\textbf {\bibinfo {volume} {101}},\ \bibinfo {pages}
  {022316} (\bibinfo {year} {2020})}\BibitemShut {NoStop}%
\bibitem [{\citenamefont {Wossnig}\ \emph {et~al.}(2018)\citenamefont
  {Wossnig}, \citenamefont {Zhao},\ and\ \citenamefont {Prakash}}]{26}%
  \BibitemOpen
  \bibfield  {author} {\bibinfo {author} {\bibfnamefont {L.}~\bibnamefont
  {Wossnig}}, \bibinfo {author} {\bibfnamefont {Z.}~\bibnamefont {Zhao}}, \
  and\ \bibinfo {author} {\bibfnamefont {A.}~\bibnamefont {Prakash}},\
  }\href@noop {} {\bibfield  {journal} {\bibinfo  {journal} {Physical review
  letters}\ }\textbf {\bibinfo {volume} {120}},\ \bibinfo {pages} {050502}
  (\bibinfo {year} {2018})}\BibitemShut {NoStop}%
\bibitem [{\citenamefont {Botsinis}\ \emph {et~al.}(2014)\citenamefont
  {Botsinis}, \citenamefont {Ng},\ and\ \citenamefont {Hanzo}}]{27}%
  \BibitemOpen
  \bibfield  {author} {\bibinfo {author} {\bibfnamefont {P.}~\bibnamefont
  {Botsinis}}, \bibinfo {author} {\bibfnamefont {S.~X.}\ \bibnamefont {Ng}}, \
  and\ \bibinfo {author} {\bibfnamefont {L.}~\bibnamefont {Hanzo}},\ }in\
  \href@noop {} {\emph {\bibinfo {booktitle} {2014 IEEE International
  Conference on Communications (ICC)}}}\ (\bibinfo {organization} {IEEE},\
  \bibinfo {year} {2014})\ pp.\ \bibinfo {pages} {5592--5597}\BibitemShut
  {NoStop}%
\bibitem [{\citenamefont {Abdullah}\ \emph {et~al.}(2018)\citenamefont
  {Abdullah}, \citenamefont {Tsimenidis},\ and\ \citenamefont {Johnston}}]{28}%
  \BibitemOpen
  \bibfield  {author} {\bibinfo {author} {\bibfnamefont {Z.}~\bibnamefont
  {Abdullah}}, \bibinfo {author} {\bibfnamefont {C.~C.}\ \bibnamefont
  {Tsimenidis}}, \ and\ \bibinfo {author} {\bibfnamefont {M.}~\bibnamefont
  {Johnston}},\ }in\ \href@noop {} {\emph {\bibinfo {booktitle} {2018 IEEE
  Wireless Communications and Networking Conference (WCNC)}}}\ (\bibinfo
  {organization} {IEEE},\ \bibinfo {year} {2018})\ pp.\ \bibinfo {pages}
  {1--6}\BibitemShut {NoStop}%
\bibitem [{\citenamefont {Alanis}\ \emph {et~al.}(2014)\citenamefont {Alanis},
  \citenamefont {Botsinis}, \citenamefont {Ng},\ and\ \citenamefont
  {Hanzo}}]{29}%
  \BibitemOpen
  \bibfield  {author} {\bibinfo {author} {\bibfnamefont {D.}~\bibnamefont
  {Alanis}}, \bibinfo {author} {\bibfnamefont {P.}~\bibnamefont {Botsinis}},
  \bibinfo {author} {\bibfnamefont {S.~X.}\ \bibnamefont {Ng}}, \ and\ \bibinfo
  {author} {\bibfnamefont {L.}~\bibnamefont {Hanzo}},\ }\href@noop {}
  {\bibfield  {journal} {\bibinfo  {journal} {IEEE Access}\ }\textbf {\bibinfo
  {volume} {2}},\ \bibinfo {pages} {614} (\bibinfo {year} {2014})}\BibitemShut
  {NoStop}%
\bibitem [{\citenamefont {Botsinis}\ \emph {et~al.}(2016)\citenamefont
  {Botsinis}, \citenamefont {Alanis}, \citenamefont {Babar}, \citenamefont
  {Nguyen}, \citenamefont {Chandra}, \citenamefont {Ng},\ and\ \citenamefont
  {Hanzo}}]{30}%
  \BibitemOpen
  \bibfield  {author} {\bibinfo {author} {\bibfnamefont {P.}~\bibnamefont
  {Botsinis}}, \bibinfo {author} {\bibfnamefont {D.}~\bibnamefont {Alanis}},
  \bibinfo {author} {\bibfnamefont {Z.}~\bibnamefont {Babar}}, \bibinfo
  {author} {\bibfnamefont {H.~V.}\ \bibnamefont {Nguyen}}, \bibinfo {author}
  {\bibfnamefont {D.}~\bibnamefont {Chandra}}, \bibinfo {author} {\bibfnamefont
  {S.~X.}\ \bibnamefont {Ng}}, \ and\ \bibinfo {author} {\bibfnamefont
  {L.}~\bibnamefont {Hanzo}},\ }\href@noop {} {\bibfield  {journal} {\bibinfo
  {journal} {IEEE Access}\ }\textbf {\bibinfo {volume} {4}},\ \bibinfo {pages}
  {7402} (\bibinfo {year} {2016})}\BibitemShut {NoStop}%
\bibitem [{\citenamefont {Meng}\ \emph {et~al.}(2020)\citenamefont {Meng},
  \citenamefont {Yu},\ and\ \citenamefont {Zhang}}]{31}%
  \BibitemOpen
  \bibfield  {author} {\bibinfo {author} {\bibfnamefont {F.-X.}\ \bibnamefont
  {Meng}}, \bibinfo {author} {\bibfnamefont {X.-T.}\ \bibnamefont {Yu}}, \ and\
  \bibinfo {author} {\bibfnamefont {Z.-C.}\ \bibnamefont {Zhang}},\ }\href@noop
  {} {\bibfield  {journal} {\bibinfo  {journal} {Physical Review A}\ }\textbf
  {\bibinfo {volume} {101}},\ \bibinfo {pages} {012334} (\bibinfo {year}
  {2020})}\BibitemShut {NoStop}%
\bibitem [{\citenamefont {Preskill}(2018)}]{32}%
  \BibitemOpen
  \bibfield  {author} {\bibinfo {author} {\bibfnamefont {J.}~\bibnamefont
  {Preskill}},\ }\href@noop {} {\bibfield  {journal} {\bibinfo  {journal}
  {Quantum}\ }\textbf {\bibinfo {volume} {2}},\ \bibinfo {pages} {79} (\bibinfo
  {year} {2018})}\BibitemShut {NoStop}%
\bibitem [{\citenamefont {McClean}\ \emph {et~al.}(2016)\citenamefont
  {McClean}, \citenamefont {Romero}, \citenamefont {Babbush},\ and\
  \citenamefont {Aspuru-Guzik}}]{33}%
  \BibitemOpen
  \bibfield  {author} {\bibinfo {author} {\bibfnamefont {J.~R.}\ \bibnamefont
  {McClean}}, \bibinfo {author} {\bibfnamefont {J.}~\bibnamefont {Romero}},
  \bibinfo {author} {\bibfnamefont {R.}~\bibnamefont {Babbush}}, \ and\
  \bibinfo {author} {\bibfnamefont {A.}~\bibnamefont {Aspuru-Guzik}},\
  }\href@noop {} {\bibfield  {journal} {\bibinfo  {journal} {New Journal of
  Physics}\ }\textbf {\bibinfo {volume} {18}},\ \bibinfo {pages} {023023}
  (\bibinfo {year} {2016})}\BibitemShut {NoStop}%
\bibitem [{\citenamefont {Kandala}\ \emph {et~al.}(2017)\citenamefont
  {Kandala}, \citenamefont {Mezzacapo}, \citenamefont {Temme}, \citenamefont
  {Takita}, \citenamefont {Brink}, \citenamefont {Chow},\ and\ \citenamefont
  {Gambetta}}]{34}%
  \BibitemOpen
  \bibfield  {author} {\bibinfo {author} {\bibfnamefont {A.}~\bibnamefont
  {Kandala}}, \bibinfo {author} {\bibfnamefont {A.}~\bibnamefont {Mezzacapo}},
  \bibinfo {author} {\bibfnamefont {K.}~\bibnamefont {Temme}}, \bibinfo
  {author} {\bibfnamefont {M.}~\bibnamefont {Takita}}, \bibinfo {author}
  {\bibfnamefont {M.}~\bibnamefont {Brink}}, \bibinfo {author} {\bibfnamefont
  {J.~M.}\ \bibnamefont {Chow}}, \ and\ \bibinfo {author} {\bibfnamefont
  {J.~M.}\ \bibnamefont {Gambetta}},\ }\href@noop {} {\bibfield  {journal}
  {\bibinfo  {journal} {Nature}\ }\textbf {\bibinfo {volume} {549}},\ \bibinfo
  {pages} {242} (\bibinfo {year} {2017})}\BibitemShut {NoStop}%
\bibitem [{\citenamefont {Cerezo}\ \emph
  {et~al.}(2020{\natexlab{a}})\citenamefont {Cerezo}, \citenamefont {Sharma},
  \citenamefont {Arrasmith},\ and\ \citenamefont {Coles}}]{35}%
  \BibitemOpen
  \bibfield  {author} {\bibinfo {author} {\bibfnamefont {M.}~\bibnamefont
  {Cerezo}}, \bibinfo {author} {\bibfnamefont {K.}~\bibnamefont {Sharma}},
  \bibinfo {author} {\bibfnamefont {A.}~\bibnamefont {Arrasmith}}, \ and\
  \bibinfo {author} {\bibfnamefont {P.~J.}\ \bibnamefont {Coles}},\ }\href@noop
  {} {\bibfield  {journal} {\bibinfo  {journal} {arXiv preprint
  arXiv:2004.01372}\ } (\bibinfo {year} {2020}{\natexlab{a}})}\BibitemShut
  {NoStop}%
\bibitem [{\citenamefont {Liu}\ \emph {et~al.}(2019)\citenamefont {Liu},
  \citenamefont {Zhang}, \citenamefont {Wan},\ and\ \citenamefont {Wang}}]{36}%
  \BibitemOpen
  \bibfield  {author} {\bibinfo {author} {\bibfnamefont {J.-G.}\ \bibnamefont
  {Liu}}, \bibinfo {author} {\bibfnamefont {Y.-H.}\ \bibnamefont {Zhang}},
  \bibinfo {author} {\bibfnamefont {Y.}~\bibnamefont {Wan}}, \ and\ \bibinfo
  {author} {\bibfnamefont {L.}~\bibnamefont {Wang}},\ }\href@noop {} {\bibfield
   {journal} {\bibinfo  {journal} {Physical Review Research}\ }\textbf
  {\bibinfo {volume} {1}},\ \bibinfo {pages} {023025} (\bibinfo {year}
  {2019})}\BibitemShut {NoStop}%
\bibitem [{\citenamefont {Higgott}\ \emph {et~al.}(2019)\citenamefont
  {Higgott}, \citenamefont {Wang},\ and\ \citenamefont {Brierley}}]{37}%
  \BibitemOpen
  \bibfield  {author} {\bibinfo {author} {\bibfnamefont {O.}~\bibnamefont
  {Higgott}}, \bibinfo {author} {\bibfnamefont {D.}~\bibnamefont {Wang}}, \
  and\ \bibinfo {author} {\bibfnamefont {S.}~\bibnamefont {Brierley}},\
  }\href@noop {} {\bibfield  {journal} {\bibinfo  {journal} {Quantum}\ }\textbf
  {\bibinfo {volume} {3}},\ \bibinfo {pages} {156} (\bibinfo {year}
  {2019})}\BibitemShut {NoStop}%
\bibitem [{\citenamefont {Jones}\ \emph {et~al.}(2019)\citenamefont {Jones},
  \citenamefont {Endo}, \citenamefont {McArdle}, \citenamefont {Yuan},\ and\
  \citenamefont {Benjamin}}]{38}%
  \BibitemOpen
  \bibfield  {author} {\bibinfo {author} {\bibfnamefont {T.}~\bibnamefont
  {Jones}}, \bibinfo {author} {\bibfnamefont {S.}~\bibnamefont {Endo}},
  \bibinfo {author} {\bibfnamefont {S.}~\bibnamefont {McArdle}}, \bibinfo
  {author} {\bibfnamefont {X.}~\bibnamefont {Yuan}}, \ and\ \bibinfo {author}
  {\bibfnamefont {S.~C.}\ \bibnamefont {Benjamin}},\ }\href@noop {} {\bibfield
  {journal} {\bibinfo  {journal} {Physical Review A}\ }\textbf {\bibinfo
  {volume} {99}},\ \bibinfo {pages} {062304} (\bibinfo {year}
  {2019})}\BibitemShut {NoStop}%
\bibitem [{\citenamefont {Wang}\ \emph
  {et~al.}(2020{\natexlab{a}})\citenamefont {Wang}, \citenamefont {Song},\ and\
  \citenamefont {Wang}}]{39}%
  \BibitemOpen
  \bibfield  {author} {\bibinfo {author} {\bibfnamefont {X.}~\bibnamefont
  {Wang}}, \bibinfo {author} {\bibfnamefont {Z.}~\bibnamefont {Song}}, \ and\
  \bibinfo {author} {\bibfnamefont {Y.}~\bibnamefont {Wang}},\ }\href@noop {}
  {\bibfield  {journal} {\bibinfo  {journal} {arXiv preprint arXiv:2006.02336}\
  } (\bibinfo {year} {2020}{\natexlab{a}})}\BibitemShut {NoStop}%
\bibitem [{\citenamefont {Bravo-Prieto}\ \emph {et~al.}(2020)\citenamefont
  {Bravo-Prieto}, \citenamefont {Garc{\'\i}a-Mart{\'\i}n},\ and\ \citenamefont
  {Latorre}}]{40}%
  \BibitemOpen
  \bibfield  {author} {\bibinfo {author} {\bibfnamefont {C.}~\bibnamefont
  {Bravo-Prieto}}, \bibinfo {author} {\bibfnamefont {D.}~\bibnamefont
  {Garc{\'\i}a-Mart{\'\i}n}}, \ and\ \bibinfo {author} {\bibfnamefont {J.~I.}\
  \bibnamefont {Latorre}},\ }\href@noop {} {\bibfield  {journal} {\bibinfo
  {journal} {Physical Review A}\ }\textbf {\bibinfo {volume} {101}},\ \bibinfo
  {pages} {062310} (\bibinfo {year} {2020})}\BibitemShut {NoStop}%
\bibitem [{\citenamefont {Huang}\ \emph {et~al.}(2019)\citenamefont {Huang},
  \citenamefont {Bharti},\ and\ \citenamefont {Rebentrost}}]{41}%
  \BibitemOpen
  \bibfield  {author} {\bibinfo {author} {\bibfnamefont {H.-Y.}\ \bibnamefont
  {Huang}}, \bibinfo {author} {\bibfnamefont {K.}~\bibnamefont {Bharti}}, \
  and\ \bibinfo {author} {\bibfnamefont {P.}~\bibnamefont {Rebentrost}},\
  }\href@noop {} {\bibfield  {journal} {\bibinfo  {journal} {arXiv preprint
  arXiv:1909.07344}\ } (\bibinfo {year} {2019})}\BibitemShut {NoStop}%
\bibitem [{\citenamefont {Xu}\ \emph {et~al.}(2019)\citenamefont {Xu},
  \citenamefont {Sun}, \citenamefont {Endo}, \citenamefont {Li}, \citenamefont
  {Benjamin},\ and\ \citenamefont {Yuan}}]{42}%
  \BibitemOpen
  \bibfield  {author} {\bibinfo {author} {\bibfnamefont {X.}~\bibnamefont
  {Xu}}, \bibinfo {author} {\bibfnamefont {J.}~\bibnamefont {Sun}}, \bibinfo
  {author} {\bibfnamefont {S.}~\bibnamefont {Endo}}, \bibinfo {author}
  {\bibfnamefont {Y.}~\bibnamefont {Li}}, \bibinfo {author} {\bibfnamefont
  {S.~C.}\ \bibnamefont {Benjamin}}, \ and\ \bibinfo {author} {\bibfnamefont
  {X.}~\bibnamefont {Yuan}},\ }\href@noop {} {\bibfield  {journal} {\bibinfo
  {journal} {arXiv preprint arXiv:1909.03898}\ } (\bibinfo {year}
  {2019})}\BibitemShut {NoStop}%
\bibitem [{\citenamefont {Bravo-Prieto}\ \emph {et~al.}(2019)\citenamefont
  {Bravo-Prieto}, \citenamefont {LaRose}, \citenamefont {Cerezo}, \citenamefont
  {Subasi}, \citenamefont {Cincio},\ and\ \citenamefont {Coles}}]{43}%
  \BibitemOpen
  \bibfield  {author} {\bibinfo {author} {\bibfnamefont {C.}~\bibnamefont
  {Bravo-Prieto}}, \bibinfo {author} {\bibfnamefont {R.}~\bibnamefont
  {LaRose}}, \bibinfo {author} {\bibfnamefont {M.}~\bibnamefont {Cerezo}},
  \bibinfo {author} {\bibfnamefont {Y.}~\bibnamefont {Subasi}}, \bibinfo
  {author} {\bibfnamefont {L.}~\bibnamefont {Cincio}}, \ and\ \bibinfo {author}
  {\bibfnamefont {P.~J.}\ \bibnamefont {Coles}},\ }\href@noop {} {\bibfield
  {journal} {\bibinfo  {journal} {arXiv preprint arXiv:1909.05820}\ } (\bibinfo
  {year} {2019})}\BibitemShut {NoStop}%
\bibitem [{\citenamefont {Cerezo}\ \emph
  {et~al.}(2020{\natexlab{b}})\citenamefont {Cerezo}, \citenamefont {Poremba},
  \citenamefont {Cincio},\ and\ \citenamefont {Coles}}]{44}%
  \BibitemOpen
  \bibfield  {author} {\bibinfo {author} {\bibfnamefont {M.}~\bibnamefont
  {Cerezo}}, \bibinfo {author} {\bibfnamefont {A.}~\bibnamefont {Poremba}},
  \bibinfo {author} {\bibfnamefont {L.}~\bibnamefont {Cincio}}, \ and\ \bibinfo
  {author} {\bibfnamefont {P.~J.}\ \bibnamefont {Coles}},\ }\href@noop {}
  {\bibfield  {journal} {\bibinfo  {journal} {Quantum}\ }\textbf {\bibinfo
  {volume} {4}},\ \bibinfo {pages} {248} (\bibinfo {year}
  {2020}{\natexlab{b}})}\BibitemShut {NoStop}%
\bibitem [{\citenamefont {Chowdhury}\ \emph {et~al.}(2020)\citenamefont
  {Chowdhury}, \citenamefont {Low},\ and\ \citenamefont {Wiebe}}]{45}%
  \BibitemOpen
  \bibfield  {author} {\bibinfo {author} {\bibfnamefont {A.~N.}\ \bibnamefont
  {Chowdhury}}, \bibinfo {author} {\bibfnamefont {G.~H.}\ \bibnamefont {Low}},
  \ and\ \bibinfo {author} {\bibfnamefont {N.}~\bibnamefont {Wiebe}},\
  }\href@noop {} {\bibfield  {journal} {\bibinfo  {journal} {arXiv preprint
  arXiv:2002.00055}\ } (\bibinfo {year} {2020})}\BibitemShut {NoStop}%
\bibitem [{\citenamefont {Wang}\ \emph
  {et~al.}(2020{\natexlab{b}})\citenamefont {Wang}, \citenamefont {Li},\ and\
  \citenamefont {Wang}}]{46}%
  \BibitemOpen
  \bibfield  {author} {\bibinfo {author} {\bibfnamefont {Y.}~\bibnamefont
  {Wang}}, \bibinfo {author} {\bibfnamefont {G.}~\bibnamefont {Li}}, \ and\
  \bibinfo {author} {\bibfnamefont {X.}~\bibnamefont {Wang}},\ }\href@noop {}
  {\bibfield  {journal} {\bibinfo  {journal} {arXiv preprint arXiv:2005.08797}\
  } (\bibinfo {year} {2020}{\natexlab{b}})}\BibitemShut {NoStop}%
\bibitem [{\citenamefont {Endo}\ \emph {et~al.}(2019)\citenamefont {Endo},
  \citenamefont {Kurata},\ and\ \citenamefont {Nakagawa}}]{47}%
  \BibitemOpen
  \bibfield  {author} {\bibinfo {author} {\bibfnamefont {S.}~\bibnamefont
  {Endo}}, \bibinfo {author} {\bibfnamefont {I.}~\bibnamefont {Kurata}}, \ and\
  \bibinfo {author} {\bibfnamefont {Y.~O.}\ \bibnamefont {Nakagawa}},\
  }\href@noop {} {\bibfield  {journal} {\bibinfo  {journal} {arXiv preprint
  arXiv:1909.12250}\ } (\bibinfo {year} {2019})}\BibitemShut {NoStop}%
\bibitem [{\citenamefont {Temme}\ \emph {et~al.}(2017)\citenamefont {Temme},
  \citenamefont {Bravyi},\ and\ \citenamefont {Gambetta}}]{48}%
  \BibitemOpen
  \bibfield  {author} {\bibinfo {author} {\bibfnamefont {K.}~\bibnamefont
  {Temme}}, \bibinfo {author} {\bibfnamefont {S.}~\bibnamefont {Bravyi}}, \
  and\ \bibinfo {author} {\bibfnamefont {J.~M.}\ \bibnamefont {Gambetta}},\
  }\href@noop {} {\bibfield  {journal} {\bibinfo  {journal} {Physical review
  letters}\ }\textbf {\bibinfo {volume} {119}},\ \bibinfo {pages} {180509}
  (\bibinfo {year} {2017})}\BibitemShut {NoStop}%
\bibitem [{\citenamefont {McArdle}\ \emph {et~al.}(2019)\citenamefont
  {McArdle}, \citenamefont {Yuan},\ and\ \citenamefont {Benjamin}}]{49}%
  \BibitemOpen
  \bibfield  {author} {\bibinfo {author} {\bibfnamefont {S.}~\bibnamefont
  {McArdle}}, \bibinfo {author} {\bibfnamefont {X.}~\bibnamefont {Yuan}}, \
  and\ \bibinfo {author} {\bibfnamefont {S.}~\bibnamefont {Benjamin}},\
  }\href@noop {} {\bibfield  {journal} {\bibinfo  {journal} {Physical review
  letters}\ }\textbf {\bibinfo {volume} {122}},\ \bibinfo {pages} {180501}
  (\bibinfo {year} {2019})}\BibitemShut {NoStop}%
\bibitem [{\citenamefont {Strikis}\ \emph {et~al.}(2020)\citenamefont
  {Strikis}, \citenamefont {Qin}, \citenamefont {Chen}, \citenamefont
  {Benjamin},\ and\ \citenamefont {Li}}]{50}%
  \BibitemOpen
  \bibfield  {author} {\bibinfo {author} {\bibfnamefont {A.}~\bibnamefont
  {Strikis}}, \bibinfo {author} {\bibfnamefont {D.}~\bibnamefont {Qin}},
  \bibinfo {author} {\bibfnamefont {Y.}~\bibnamefont {Chen}}, \bibinfo {author}
  {\bibfnamefont {S.~C.}\ \bibnamefont {Benjamin}}, \ and\ \bibinfo {author}
  {\bibfnamefont {Y.}~\bibnamefont {Li}},\ }\href@noop {} {\bibfield  {journal}
  {\bibinfo  {journal} {arXiv preprint arXiv:2005.07601}\ } (\bibinfo {year}
  {2020})}\BibitemShut {NoStop}%
\bibitem [{\citenamefont {Nakanishi}\ \emph {et~al.}(2019)\citenamefont
  {Nakanishi}, \citenamefont {Mitarai},\ and\ \citenamefont {Fujii}}]{51}%
  \BibitemOpen
  \bibfield  {author} {\bibinfo {author} {\bibfnamefont {K.~M.}\ \bibnamefont
  {Nakanishi}}, \bibinfo {author} {\bibfnamefont {K.}~\bibnamefont {Mitarai}},
  \ and\ \bibinfo {author} {\bibfnamefont {K.}~\bibnamefont {Fujii}},\
  }\href@noop {} {\bibfield  {journal} {\bibinfo  {journal} {Physical Review
  Research}\ }\textbf {\bibinfo {volume} {1}},\ \bibinfo {pages} {033062}
  (\bibinfo {year} {2019})}\BibitemShut {NoStop}%
\bibitem [{\citenamefont {Marcello}\ \emph {et~al.}(2019)\citenamefont
  {Marcello}, \citenamefont {Delfina}, \citenamefont {Perdomo}, \citenamefont
  {Leyton-Ortega}, \citenamefont {Yunseong},\ and\ \citenamefont
  {Perdomo-Ortiz}}]{52}%
  \BibitemOpen
  \bibfield  {author} {\bibinfo {author} {\bibfnamefont {B.}~\bibnamefont
  {Marcello}}, \bibinfo {author} {\bibfnamefont {G.-P.}\ \bibnamefont
  {Delfina}}, \bibinfo {author} {\bibfnamefont {O.}~\bibnamefont {Perdomo}},
  \bibinfo {author} {\bibfnamefont {V.}~\bibnamefont {Leyton-Ortega}}, \bibinfo
  {author} {\bibfnamefont {N.}~\bibnamefont {Yunseong}}, \ and\ \bibinfo
  {author} {\bibfnamefont {A.}~\bibnamefont {Perdomo-Ortiz}},\ }\href@noop {}
  {\bibfield  {journal} {\bibinfo  {journal} {NPJ Quantum Information}\
  }\textbf {\bibinfo {volume} {5}} (\bibinfo {year} {2019})}\BibitemShut
  {NoStop}%
\bibitem [{\citenamefont {Killoran}\ \emph {et~al.}(2019)\citenamefont
  {Killoran}, \citenamefont {Bromley}, \citenamefont {Arrazola}, \citenamefont
  {Schuld}, \citenamefont {Quesada},\ and\ \citenamefont {Lloyd}}]{53}%
  \BibitemOpen
  \bibfield  {author} {\bibinfo {author} {\bibfnamefont {N.}~\bibnamefont
  {Killoran}}, \bibinfo {author} {\bibfnamefont {T.~R.}\ \bibnamefont
  {Bromley}}, \bibinfo {author} {\bibfnamefont {J.~M.}\ \bibnamefont
  {Arrazola}}, \bibinfo {author} {\bibfnamefont {M.}~\bibnamefont {Schuld}},
  \bibinfo {author} {\bibfnamefont {N.}~\bibnamefont {Quesada}}, \ and\
  \bibinfo {author} {\bibfnamefont {S.}~\bibnamefont {Lloyd}},\ }\href@noop {}
  {\bibfield  {journal} {\bibinfo  {journal} {Physical Review Research}\
  }\textbf {\bibinfo {volume} {1}},\ \bibinfo {pages} {033063} (\bibinfo {year}
  {2019})}\BibitemShut {NoStop}%
\bibitem [{\citenamefont {Perdomo}\ \emph {et~al.}(2019)\citenamefont
  {Perdomo}, \citenamefont {Leyton-Ortega},\ and\ \citenamefont
  {Perdomo}}]{54}%
  \BibitemOpen
  \bibfield  {author} {\bibinfo {author} {\bibfnamefont {A.}~\bibnamefont
  {Perdomo}}, \bibinfo {author} {\bibfnamefont {V.}~\bibnamefont
  {Leyton-Ortega}}, \ and\ \bibinfo {author} {\bibfnamefont {O.}~\bibnamefont
  {Perdomo}},\ }\href@noop {} {\bibfield  {journal} {\bibinfo  {journal} {APS}\
  }\textbf {\bibinfo {volume} {2019}},\ \bibinfo {pages} {C42} (\bibinfo {year}
  {2019})}\BibitemShut {NoStop}%
\bibitem [{\citenamefont {Brassard}\ \emph {et~al.}(2002)\citenamefont
  {Brassard}, \citenamefont {Hoyer}, \citenamefont {Mosca},\ and\ \citenamefont
  {Tapp}}]{55}%
  \BibitemOpen
  \bibfield  {author} {\bibinfo {author} {\bibfnamefont {G.}~\bibnamefont
  {Brassard}}, \bibinfo {author} {\bibfnamefont {P.}~\bibnamefont {Hoyer}},
  \bibinfo {author} {\bibfnamefont {M.}~\bibnamefont {Mosca}}, \ and\ \bibinfo
  {author} {\bibfnamefont {A.}~\bibnamefont {Tapp}},\ }\href@noop {} {\bibfield
   {journal} {\bibinfo  {journal} {Contemporary Mathematics}\ }\textbf
  {\bibinfo {volume} {305}},\ \bibinfo {pages} {53} (\bibinfo {year}
  {2002})}\BibitemShut {NoStop}%
\end{thebibliography}%

\end{document}